\documentclass[11pt]{llncs}

\usepackage{fullpage} 
\usepackage{amsmath}
\usepackage{amsfonts}
\usepackage{amssymb}
\usepackage{graphicx}
\usepackage{array}
\usepackage{tikz}
\usepackage{float}
\usepackage{sectsty}
\usepackage{booktabs} 
\usepackage{multirow}
\usepackage{soul} 
\usepackage{microtype}
\usepackage{algorithmic}
\usepackage[lined,boxed,ruled,commentsnumbered]{algorithm2e}
\usepackage[absolute]{textpos}

\newcommand{\mc}{\mathcal}
\newcommand{\PP}{$\textsf{P}$}

\newcommand{\WO}{$\textsf{W[1]}$}
\newcommand{\NP}{$\textsf{NP}$}
\newcommand{\FPT}{$\textsf{FPT}$}
\newcommand{\PSPACE}{$\textsf{PSPACE}$}
\newcommand{\Oh}{\mathcal{O}}

\makeatletter
\newcommand\xleftrightarrow[2][]{%
  \ext@arrow 9999{\longleftrightarrowfill@}{#1}{#2}}
\newcommand\longleftrightarrowfill@{%
  \arrowfill@\leftarrow\relbar\rightarrow}
\makeatother

\newtheorem{Theorem}{Theorem}
\newtheorem{Proposition}[Theorem]{Proposition}
\newtheorem{Lemma}[Theorem]{Lemma} 
\newtheorem{Definition}[Theorem]{Definition}
\numberwithin{Theorem}{subsection}

\begin{document}
\title{Vertex Cover Reconfiguration and Beyond}
\author
{
    Amer E. Mouawad\inst{1} \and
    Naomi Nishimura\inst{2}\thanks{Research supported by the Natural Science and Engineering Research Council of Canada.} \and
    Venkatesh Raman\inst{3} \and
    Sebastian Siebertz\inst{4}\thanks{The work of Sebastian Siebertz is supported by the National Science Centre of Poland via POLONEZ 
  Grant Agreement UMO-2015/19/P/ST6/03998, 
which has received funding from the European Union's Horizon 2020 research and 
innovation programme (Marie Sk\l odowska-Curie Grant Agreement No.\ 665778).}
}
\institute
{
    University of Bergen, Norway.\\
    \email{a.mouawad@uib.no}
    \and
    University of Waterloo, Ontario, Canada.\\
    \email{nishi@uwaterloo.ca}
    \and
    The Institute of Mathematical Sciences, Chennai, India.\\
    \email{vraman@imsc.res.in}
    \and
    Institute of Informatics, University of Warsaw, Poland.\\
    \email{siebertz@mimuw.edu.pl}
}
\maketitle

\begin{textblock}{5}(12.7,13.9)
\includegraphics[width=38px]{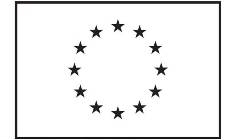}
\end{textblock}

\sloppy

\begin{abstract} 
In the {Vertex Cover Reconfiguration} (VCR) problem, given a graph $G$, positive integers $k$ and $\ell$ and two vertex covers $S$
and $T$ of $G$ of size at most $k$, we determine whether $S$ can be
transformed into $T$ by a sequence of at most $\ell$ vertex additions
or removals such that every operation results in a vertex cover
of size at most $k$. Motivated by results establishing the
\WO-hardness of VCR 
when parameterized by $\ell$, we delineate the complexity of 
the problem restricted to various graph classes. In particular, 
we show that VCR remains \WO-hard on bipartite graphs, is \NP-hard, 
but fixed-parameter tractable on (regular) graphs of bounded degree and more generally on nowhere dense graphs 
and is solvable in polynomial time on trees and (with some additional restrictions) on cactus graphs. 
\end{abstract}

\section{Introduction}\label{section-introduction}
Under the {reconfiguration framework}, we consider structural and
algorithmic questions related to the solution space of a search
problem ${\cal Q}$. Given an instance ${\cal I}$, an optional range $[r_l, r_u]$
bounding a numerically-quantifiable property $\Psi$ of
feasible solutions for ${\cal Q}$ and a symmetric adjacency relation
(usually polynomially-testable) ${\cal A}$ on the set of
feasible solutions, we can construct a
{reconfiguration graph} $\mc{R}_{\cal Q}({\cal I}, r_l, r_u)$ for each
instance ${\cal I}$ of ${\cal Q}$. The nodes of
$\mc{R}_{\cal Q}({\cal I}, r_l, r_u)$ correspond to the feasible solutions of ${\cal I}$
having $r_l \leq \Psi \leq r_u$, and there is an edge between two
nodes whenever the corresponding solutions are adjacent under
${\cal A}$. An edge can be seen as a 
{reconfiguration step} transforming one solution into the other.
Given two feasible solutions for ${\cal I}$, $S$
and $T$, one can ask if there exists a walk ({reconfiguration sequence})
in $R_{\cal Q}({\cal I}, r_l, r_u)$ from
$S$ to $T$, or for the shortest such walk. On the structural side,
one can ask about the diameter of reconfiguration graph
$\mc{R}_{\cal Q}({\cal I}, r_l, r_u)$ or whether it
is connected with respect to some or any ${\cal I}$, fixed
${\cal A}$ and fixed $\Psi$.

These types of reconfiguration questions have received considerable attention
in recent years~\mbox{\cite{FHHH11,GKMP09,IDHPSUU11,IKOZ12,KMM11}} and are
interesting for a variety of reasons. From an algorithmic standpoint, reconfiguration
problems model dynamic situations in which we seek to transform a solution into
a more desirable one, maintaining feasibility during the process. Reconfiguration
also models questions of evolution; it can represent the evolution of a genotype
where only individual mutations are allowed and all genotypes must satisfy a
certain fitness threshold, i.e., be feasible.
Moreover, the study of reconfiguration yields insights into the
structure of the solution space of the underlying problem, crucial for the design of efficient
algorithms. In fact, one of the initial motivations behind such questions was to study
the performance of heuristics \cite{GKMP09} and random sampling methods \cite{CVJ08},
where connectivity and other properties of the solution space play a 
crucial role.

Reconfiguration problems have been studied mainly under
classical complexity assumptions, with most work devoted to
determining the existence of a reconfiguration sequence between two
given solutions. For most \NP-complete problems, this
question has been shown to be \PSPACE
-complete~\cite{IDHPSUU11,IKD12,KMM12}, while for some problems in \PP,
the reconfiguration question could be either in \PP~\cite{IDHPSUU11} or 
\PSPACE-complete \cite{B12}. As \PSPACE-completeness implies that the number of
vertices in reconfiguration graphs, and therefore the length of
reconfiguration sequences, can be superpolynomial in the number of
vertices in the input graph, it is natural to ask
whether we can achieve tractability if 
we restrict the length of the sequence or other properties of the problem 
to a fixed constant. 
These results motivated Mouawad
et al.~\cite{MNRSS13} to study reconfiguration under the
{parameterized complexity} framework~\cite{DF97,DowneyF13}.

The {Vertex Cover Reconfiguration} (VCR) 
problem was shown to be fixed-parameter tractable when
parameterized by $k$ and \WO-hard when parameterized by
$\ell$~\cite{MNRSS13}; in $\mc{R}_{\textsf{VC}}(G, 0, k)$, each feasible solution for
instance $G$ is a vertex cover of size at most $k$ (a subset $S
\subseteq V(G)$ such that each edge of the graph has at least one endpoint
in $S$) and two solutions are adjacent if one can be obtained from the
other by the addition or removal of a single vertex of $G$. Motivated
by these results, we embark on a systematic investigation of the
parameterized complexity of the problem restricted to various graph~classes.

In Section \ref{section-hardness}, we start by showing that
the VCR problem parameterized by $\ell$ remains
\WO-hard when restricted to bipartite graphs. 
To obtain this result, we introduce the {$(t,d)$-bipartite constrained crown}
problem and show that it plays a central role
for determining the complexity of the reconfiguration problem.
As the {vertex cover} is solvable in polynomial time
on bipartite graphs, this result provides an example of
a search problem in \PP\ whose reconfiguration version
is \WO-hard parameterized by $\ell$, answering a
question left open by Mouawad et al.~\cite{MNRSS13}.
In Section~\ref{section-poly}, we~characterize instances
of the VCR problem solvable in time polynomial in $|V(G)|$
and apply this characterization to trees, graphs with no even cycles and 
(with some additional restrictions) to cactus graphs (we incorrectly claimed to have proved the result for cactus graphs in its full generality in an earlier version of this paper~\cite{DBLP:conf/isaac/MouawadNR14}). 
We note that a polynomial-time algorithm for even-hole-free graphs was also independently 
obtained by Kami\'{n}ski et al.~\cite{KMM12} for solving several variants of
the closely-related {independent set reconfiguration} problem. 
Moreover, VCR is known to be \PSPACE-complete on graphs of bounded treewidth~\cite{Wrochna18} (for some constant value of treewidth), but it remains open 
whether the problem is \PSPACE-complete already for graphs of treewidth at most two, or even outerplanar graphs. Our result on cactus graphs 
is a first step towards settling these questions. 
In Section~\ref{section-fpt}, we~present
the first fixed-parameter tractable algorithm for VCR parameterized by $\ell$ on graphs of bounded degree
after establishing the \NP-hardness of the problem on four-regular graphs. Finally, we~show using completely different techniques, 
and at the cost of a much worse running time, that~VCR, as well as a host of other reconfiguration problems are 
fixed-parameter tractable on nowhere dense classes of~graphs.

\section{Preliminaries}\label{section-preliminaries}
For general graph theoretic definitions, we refer
the reader to the book of Diestel \cite{D05}.
Unless otherwise stated, we assume that each graph $G$ is a
simple undirected graph with vertex set $V(G)$ and
edge set $E(G)$, where $|V(G)| = n$ and $|E(G)| = m$.
The {open neighbourhood} of a vertex $v$ is denoted by $N_G(v) = \{u \mid uv \in E(G)\}$ and the
{closed neighbourhood} by $N_G[v] = N_G(v) \cup \{v\}$.
For a set of vertices $A \subseteq V(G)$,
we define $N_G(A) = \{v \not\in A \mid uv \in E(G), u \in A\}$ and $N_G[A] = N_G(A) \cup A$.
The subgraph of $G$ induced by $A$ is denoted by $G[A]$, where
$G[A]$ has vertex set $A$ and edge set $\{uv \in E(G) \mid u, v \in A\}$.

A {walk} of length $q$ from $v_0$ to $v_q$ in $G$ is a vertex
sequence $v_0, \ldots, v_q$ such that, for all $i \in \mbox{\{0, \ldots,
$q-1$\}}$, $v_iv_{i + 1} \in E(G)$. It is a {path} if all
vertices are distinct and a {cycle} if $q \geq 3$, $v_0 = v_q$,
and $v_0, \ldots, v_{q - 1}$ is a path. 
A {matching} ${\cal{M}}(G)$ on a graph $G$ is a set
of edges of $G$ such that no two edges share a vertex;
we use $V({\cal{M}}(G))$ to denote the set of vertices 
incident to edges in ${\cal{M}}(G)$.
A set of vertices $A \subseteq V(G)$ is said to be 
{saturated} by ${\cal{M}}(G)$ if $A \subseteq V({\cal{M}}(G))$.

To avoid confusion, we refer to {nodes} in
reconfiguration graphs, as distinguished from {vertices} in the input graph.
We denote an instance of the {vertex cover reconfiguration} problem by $(G,S,T,k,\ell)$,
where $G$ is the input graph, $S$ and $T$ are the {source} and {target} vertex covers,
respectively, $k$ is the {maximum allowed capacity} and $\ell$ is an upper
bound on the length of the reconfiguration sequence we seek. 
By~a slight abuse of notation, we use upper case letters to refer
to both a node in the reconfiguration
graph, as well as the corresponding vertex cover.
For any node $S \in V(\mc{R}_{\textsf{VC}}(G, 0, k))$,
the quantity $k - |S|$ corresponds to the {available capacity} at $S$.
We partition $V(G)$ into the sets
$C_{ST} = S \cap T$ (vertices common to $S$ and $T$), $S_R = S \setminus C_{ST}$
(vertices to be removed from $S$ in the course of reconfiguration),
$T_A = T \setminus C_{ST}$ (vertices to be added to form $T$) and $O_{ST} =
V(G) \setminus (S \cup T) = V(G) \setminus (C_{ST} \cup S_R \cup T_A)$ (all
other vertices).
To simplify notation, we sometimes use $G_{\Delta}$ to denote the graph induced
by the vertices in the symmetric difference of
$S$ and $T$, i.e., $G_{\Delta} = G[S \Delta T] = G[S_R \cup T_A]$.
We say a vertex is {touched} in the course of a reconfiguration
sequence from $S$ to $T$ if $v$ is either added or removed at least once.
We say a vertex $v$, in a vertex cover
$S$, is {removable} if and only if $v \in S$ and $N_G(v) \subseteq S$.

\begin{Proposition}\label{obs-bipartite}
For any graph $G$ and any two vertex covers $S$ and $T$ of $G$,
$G_{\Delta} = G[S_R \cup T_A]$ is bipartite. Moreover,
there are no edges between vertices in $S_R \cup T_A$ and
vertices in $O_{ST}$.
\end{Proposition}

\begin{proof}
None of the vertices in $S_R$ are included in $T$. Since $T$ is a vertex cover of $G$,
each edge of $G$ must have an endpoint in $T$, and hence, $G[S_R]$ must be an independent set.
Similar arguments apply to $G[T_A]$ and to show that there
are no edges between vertices in $S_R \cup T_A$ and vertices in $O_{ST}$.
\end{proof}

\begin{Proposition}\label{touch-times}
For a graph $G$ and any two vertex covers $S$ and $T$ of $G$,
any vertex in $S_R \cup T_A$ must be touched an odd number of times
and any vertex not in $S_R \cup T_A$ must be touched an even number of times
in any reconfiguration sequence of length at most $\ell$ from $S$ to $T$.
Moreover, any vertex can be touched at most $\ell - |S_R \cup T_A| + 1$ times.
\end{Proposition}

Throughout this work, we implicitly
consider the {vertex cover reconfiguration} problem
as a parameterized problem with $\ell$ as the parameter.
The reader is referred to the books of Downey and 
Fellows for more on parameterized complexity~\cite{DF97,DowneyF13}.

\section{Representing Reconfiguration Sequences}\label{section-represent}
There are multiple ways of representing a reconfiguration sequence
between two vertex covers of a graph $G$.
In Sections~\ref{section-hardness} and~\ref{section-poly},
we focus on a representation that consists of an
ordered sequence of vertex covers or nodes
in the reconfiguration graph.
Given a graph $G$ and two vertex covers of $G$, $A_0$~and $A_j$,
we denote a reconfiguration sequence from $A_0$ to $A_j$ by $\alpha = (A_0, A_1, \ldots, A_j)$,
where $A_i$ is a vertex cover of $G$ and $A_{i}$ is obtained from $A_{i - 1}$
by either the removal or the addition of a single vertex from $A_{i - 1}$ for all $0 < i \leq j$.
For any pair of consecutive vertex covers ($A_{i - 1}$, $A_{i}$) in $\alpha$, we say
$A_{i}$ ($A_{i - 1}$) is the {successor} ({predecessor}) of $A_{i - 1}$ ($A_{i}$).
A reconfiguration sequence $\beta = (A_0, A_1, \ldots, A_i)$
is a {prefix} of $\alpha = (A_0, A_1, \ldots, A_j)$ if $i < j$.

In Section~\ref{section-fpt}, we use the notion of edit sequences. 
We assume all vertices of $G$ are labelled from one to $n$, 
i.e., $V(G) = \{v_1, v_2, \ldots, v_n\}$. 
We let ${\cal E}_a = \{a_1, \ldots, a_n\}$ and ${\cal E}_r = \{r_1, \ldots, r_n\}$ denote the
sets of {addition markers} and {removal markers}, respectively. 
An {edit sequence} $\alpha$ is an ordered sequence of
elements obtained from the full set of {markers}
${\cal E} = {\cal E}_a \cup {\cal E}_r$,
where $a_i$ stands for the addition of vertex~$v_i$, 
$r_j$ stands for the removal of vertex $v_j$ and $1 \leq i,j \leq n$. 
The length of $\alpha$, $|\alpha|$, is equal to the total number of markers in $\alpha$. 
We let $\alpha[p] \in {\cal E}$, $1 \leq p \leq |\alpha|$, denote
the marker at position $p$ in $\alpha$. 
We say $\beta$ is a {segment} of $\alpha$ whenever $\beta$
consists of a subsequence of $\alpha$ with no gaps.
The length of a segment is defined as the total
number of markers it contains. We use the notation
$\alpha[p_1, p_2]$, $1 \leq p_1,\ p_2 \leq |\alpha|$, to denote
the segment starting at position $p_1$ and ending at position $p_2$.
Two segments $\beta$ and $\beta'$ are {consecutive} whenever $\beta'$ occurs later
than $\beta$ in $\alpha$ and there are no gaps between $\beta$ and $\beta'$. 
For any pair of consecutive segments $\beta$
and $\beta'$ in $\alpha$, we say $\beta'$ ($\beta$) is
the {successor} ({predecessor}) of $\beta$ ($\beta'$).
Given an edit sequence $\alpha$, a segment $\beta$ of $\alpha$ is a {maximal addition segment} if 
$\beta$ is a maximal subsequence of $\alpha$ consisting of only addition markers and no gaps. 
Similarly, $\beta$ is a {maximal removal segment} if 
$\beta$ is a maximal subsequence of $\alpha$ consisting of only removal markers and no gaps. 

We now discuss how edit sequences relate to reconfiguration sequences.
Given a graph $G$ and an edit sequence $\alpha$,
we use $V(\alpha)$ to denote the set of vertices touched in
$\alpha$, i.e., $V(\alpha) = \{v_i \mid a_i \in \alpha \vee r_i \in \alpha\}$.
We let $V(S, \alpha)$ denote the 
set of vertices obtained after executing all reconfiguration steps in
$\alpha$ on $G$ starting from some vertex cover $S$ of $G$. 
We say $\alpha$ is {valid} whenever every set
$V(S, \alpha[1, p])$, $1 \leq p \leq |\alpha|$, is a vertex cover of $G$,
and we say $\alpha$ is {invalid} otherwise.
Note that even if $|S| \leq k$, $\alpha$ is not necessarily a walk in the reconfiguration graph 
$\mc{R}_{\textsf{VC}}(G, 0, k)$, as $\alpha$ might violate the maximum
allowed capacity constraint $k$. Hence, we let
$\textsf{cap}(\alpha) = \textsf{max}_{1 \leq p \leq |\alpha|}(|V(S, \alpha[1, p])|)$, and we
say $\alpha$ is {tight} whenever it is valid and $\textsf{cap}(\alpha) \leq k$. 

\begin{Proposition}
Given a graph $G$ and two vertex covers $S$ and $T$ of $G$,
an edit sequence $\alpha$ is a reconfiguration sequence from $S$ to $T$
if and only if $\alpha$ is a tight edit sequence from $S$ to $T$.
\end{Proposition}

\section{Hardness Results}\label{section-hardness}
In earlier work establishing the \WO-hardness
of the VCR problem parameterized
by $\ell$ on general graphs, it was also shown that the problem
becomes fixed-parameter tractable
whenever $\ell = |S_R \cup T_A|$~\cite{MNRSS13} (as we know exactly which vertices have to be touched, it is only a question of finding the right order of additions and removals). 
When $|S_R \cup T_A| = n$, we know from Proposition~\ref{touch-times} 
that $\ell \geq n$, since every 
vertex in $S_R \cup T_A$ must be touched at least once. 
Moreover, Proposition~\ref{obs-bipartite} implies that whenever
$|S_R \cup T_A| = n$, the input graph must be bipartite.
It is thus natural to ask
about the complexity of the problem when $\ell < n$ 
and the input graph is restricted to be bipartite.
Since the {vertex cover} problem is known to be solvable
in time polynomial in $n$ on bipartite graphs, our result is, to the best of our
knowledge, the first example of a problem solvable in polynomial time
whose reconfiguration version is \WO-hard.

For a graph $G$, a {crown} is a pair
$(W, H)$ satisfying the following properties: (i) $W \neq \emptyset$ is
an independent set of $G$; (ii) $N_G(W) = H$; and (iii) there exists
a matching in $G[W \cup H]$ that saturates $H$~\cite{ACFLSS09,CFJ04}.
$H$ is called the {head} of the crown, and the {width} of
the crown is $|H|$. Crown structures have played a central role
in the development of efficient kernelization algorithms for the
{vertex cover} problem~\cite{ACFLSS09,CFJ04}.
We define the closely-related notion of $(t,d)$-constrained crowns
and show in the remainder of this section
that the complexity of finding such structures in a bipartite graph is
central for determining the complexity of the reconfiguration problem.

We define a {$(t,d)$-constrained crown}
as a pair $(W, H)$ satisfying all properties of a regular
crown with the additional constraints that $|H| \leq t$ and
$|W| - |H| \geq d \geq 0$.
We are now ready to introduce the {$(t,d)$-Bipartite Constrained Crown}
problem, or $(t,d)$-BCC, which is formally defined as follows:

\vspace{8pt}
\begin{center}
\begin{tabular}{ll}
\toprule
\multicolumn{2}{l}{{$(t,d)$-bipartite constrained crown}}\\
{\bf Input}: & A bipartite graph $G=(A \cup B,E)$ and two positive integers $t$ and $d$\\
{\bf Parameters}: & $t$ and $d$\\
{\bf Question}: & Does $G$ have a $(t,d)$-constrained crown $(W, H)$ such that $W \subseteq A$ and $H \subseteq B$?\\
\bottomrule
\end{tabular}
\end{center}
\vspace{8pt}

\begin{Lemma}\label{lemma-bipartite-crown}
{The $(t,d)$-bipartite constrained crown} is \WO-hard
even when the input graph, $G = (A \cup B, E)$, is~$C_4$-free
and all vertices in $A$ have degree at most two.
\end{Lemma}

\begin{proof}
We give a reduction from the {$k$-clique}, known to
be \WO-hard, to the {$(k,{k \choose 2})$-bipartite constrained crown}.
For $(G, k)$ an instance of {$k$-clique}, we let
$V(G) = \{v_1, \ldots, v_n\}$ and $E(G) = \{e_1, \ldots, e_m\}$.

We first form a bipartite graph $G' = ((X \cup Z) \cup Y, E_1 \cup
E_2)$, where vertex sets $X$ and $Y$ contain one vertex for each
vertex in $V(G)$ and $Z$ contains one vertex for each edge in $E(G)$.
More formally, we set $X = \{x_1, \ldots, x_n\}$, $Y = \{y_1, \ldots,
y_n\}$, and $Z = \{z_1, \ldots, z_m\}$. The edges in $E_1$
join each pair of vertices $x_i$ and $y_i$ for $1 \le i \le n$
and the edges in $E_2$ join each vertex $z$ in $Z$ to the two vertices $y_i$ and $y_j$
corresponding to the endpoints of the edge in $E(G)$ to which $z$ corresponds.
Since each edge either joins vertices in $X$ and $Y$ or vertices in
$Y$ and $Z$, it is not difficult to see that the vertex sets $X \cup
Z$ and $Y$ form a bipartition.

By our construction, $G'$ is $C_4$-free; vertices in $X$ have degree
one, and since there are no double edges in $G$, i.e., two edges between the same pair of vertices,
no pair of vertices in $Y$ can have more than one common neighbour in $Z$. 
For $(G', k, {k \choose 2})$ an instance of $(k,{k \choose 2})$-BCC, $A = X \cup Z$ and $B = Y$,
we claim that $G$ has a clique of size $k$ if
and only if $G'$ has a $(k, {k \choose 2})$-constrained crown
$(W, H)$ such that $W \subseteq A$ and $H \subseteq B$.

If $G$ has a clique $K$ of size $k$, we set $H = \{y_i \mid v_i \in
V(K)\}$, namely the vertices in $Y$ corresponding to the vertices in the
clique. To form $W$, we choose $\{x_i \mid v_i \in V(K)\} \cup
\{z_i \mid e_i \in E(K)\}$, that is the~vertices in $X$
corresponding to the vertices in the clique and the vertices in $Z$
corresponding to the edges in the clique. Clearly, $H$ is a subset of
size $k$ of $B$, and $W$ is a subset of size $k + {k \choose 2}$ of
$A$; this implies that $|W|-|H| \geq d = {k \choose 2}$, as required.
To see why $N_{G'}(W) = H$, it suffices to note
that every vertex $x_i \in W$ is connected to exactly one
vertex $y_i \in H$, and every degree-two vertex $z_i \in W$ corresponds
to an edge in $K$ whose endpoints $v_iv_j$
must have corresponding vertices in $H$.
Moreover, due to $E_1$, there is a matching between the vertices of $H$
and the vertices of $W$ in $X$ and, hence, a matching in $G'[W \cup H]$
that saturates $H$.

We now assume that $G'$ has a $(k, {k \choose 2})$-constrained crown $(W, H)$
such that $W \subseteq X \cup Z$ and $H \subseteq Y$.
It suffices to show that $|H|$ must be equal to $k$,
$|W \cap Z|$ must be equal to ${k \choose 2}$ and,
hence, $|W \cap X|$ must be equal to $k$;
from this, we can conclude that the vertices in $\{v_i \mid y_i \in H\}$
form a clique of size $k$ in $G$ as $|W \cap Z| = {k \choose 2}$,
requiring that edges exist between each pair of vertices
in the set $\{v_i \mid y_i \in H\}$. Moreover, since
$|W \cap X| = k$ and $N_{G'}(W) = H$, a matching that saturates
$H$ can be easily found by simply picking all edges $x_iy_i$
for $y_i \in H$.

To prove the sizes of $H$ and $W$, we first observe that since
$|H| \le k$, $N_{G'}(W) = H$, and each vertex in $Y$ has exactly
one neighbour in $X$, we know that $|W \cap X| \leq |H| \leq k$.
Moreover, since $|W| = |W \cap X| + |W \cap Z|$ and $|W| - |H| \ge {k \choose 2}$,
we know that $|W \cap Z| = |W| - |W \cap X| \geq {k \choose 2} + |H| - |W \cap X| \geq {k \choose 2}$.
If $|W \cap Z| = {k \choose 2}$, our proof is complete, since by
our construction of $G'$, $H$ is a set of at most $k$ vertices in the
original graph $G$, and the subgraph induced by those vertices in $G$ has ${k \choose 2}$
edges. Hence, $|H|$ must be equal to $k$.
Suppose instead that $|W \cap Z| > {k \choose 2}$. In this case,
since each vertex of $Z$ has degree two, the number of neighbours of
$W \cap Z$ in $Y$ is greater than $k$, violating the assumptions that
$N_{G'}(W) = H$ and $|H| \le k$.
\end{proof}

We can now show the main result of this section:

\begin{Theorem}
VCR parameterized by $\ell$
and restricted to bipartite graphs is \WO-hard.
\end{Theorem}

\begin{proof}
We give a reduction from the {$(t,d)$-bipartite constrained crown}
to {vertex cover reconfiguration} in bipartite graphs.
For $(G = (A \cup B, E), t, d)$, an instance of the
 {$(t,d)$-bipartite constrained crown}, $A = \{a_1, \ldots,
a_{|A|}\}$ and $B = \{b_1, \ldots, b_{|B|}\}$, we form $G' = (X \cup Y
\cup U \cup V, E_1 \cup E_2)$ such that $X$ and $Y$ correspond to the
vertex sets $A$ and $B$, $E_1$ connects vertices in $X$ and $Y$
corresponding to vertices in $A$ and $B$ joined by edges in $G$ and
$U$, $V$ and $E_2$ form a complete bipartite graph $K_{d+t,d+t}$. More
formally, $X = \{x_1, \ldots, x_{|A|}\}$, $Y = \{y_1, \ldots,
y_{|B|}\}$, $U = \{u_1, \ldots, u_{d + t}\}$, $V = \{v_1, \ldots, v_{d +
 t}\}$, $E_1 = \{x_iy_j \mid a_ib_j \in E(G)\}$ and $E_2 =
\{u_iv_j \mid 1 \leq i \leq d + t, 1 \leq j \leq d + t\}$. 

We let $(G', S, T, k = |A| + d + 2t, \ell = 4d + 6t)$ be an instance of VCR, where $S = X \cup U$ and $T = X \cup V$.
Clearly, $|S| = |T| = |A| + d + t$. 
We claim that $G$ has a $(k,d)$-constrained crown $(W, H)$
such that $W \subseteq A$ and $H \subseteq B$ if and only if
there is a path of length at most $4d + 6t$ from $S$ to $T$.

If $G$ has such a pair $(W, H)$, we form a reconfiguration sequence
of length at most $4d + 6t$ as~follows:
\begin{enumerate}
\item[(1)] Add each vertex $y_i$ such that $b_i \in H$.
The resulting vertex cover size is $|A| + d + t + |H|$.
\item[(2)] Remove $d + |H|$ vertices $x_i$ such that $a_i \in W$.
The resulting vertex cover size is $|A| + t$.
\item[(3)] Add each vertex from $V$.
The resulting vertex cover size is $|A| + d + 2t$.
\item[(4)] Remove each vertex from $U$.
The resulting vertex cover size is $|A| + t$.
\item[(5)] Add each vertex removed in Phase 2.
The resulting vertex cover size is $|A| + d + t + |H|$.
\item[(6)] Remove each vertex added in Phase 1.
The resulting vertex cover size is $|A| + d + t$.
\end{enumerate}

The length of the sequence follows from the fact that $|H| \le t$: Phases 1 and 6 consist
of at most $t$ steps each and Phases 2, 3, 4 and 5 of at most $d + t$ steps each.
The fact that each set forms a vertex
cover is a consequence of the fact that $N_G(W) = H$.

For the converse, we observe that before removing any
vertex $u_i$, $1 \leq i \leq d + t$, from $U$, we first need to add
all $d + t$ vertices from $V$. Therefore, if there is a path of length at most $4d + 6t$ from $S$ to $T$, then
we can assume without loss of generality that there
exists a node $Q$ (i.e., a vertex cover) along this path such that:

\begin{center}
$|Q| \leq |A| + t$ and,\\
\end{center}
all vertices that were touched in order to reach node $Q$ belong to $X \cup Y$.

In other words, at node $Q$, the available capacity is greater than or equal to $d + t$,
and all edges in $G[U \cup V]$ are still covered by $U$.
We let $Q_{{\textsc{IN}}} = Q \setminus S$ and $Q_{{\textsc{OUT}}} = S \setminus Q$.
Since $S = X \cup U$, $Q_{{\textsc{IN}}} \subseteq Y$ and $Q_{{\textsc{OUT}}} \subseteq X$.
Moreover, since $|Q| = |S| + |Q_{{\textsc{IN}}}| - |Q_{{\textsc{OUT}}}| =
|A| + d + t + |Q_{{\textsc{IN}}}| - |Q_{{\textsc{OUT}}}| \leq |A| + t$,
we know that $|Q_{{\textsc{OUT}}}| - |Q_{{\textsc{IN}}}|$ must be greater than or equal to $d$.
Given that $\ell \leq 4d + 6t$ and we need exactly $2d + 2t$ steps to
add all vertices in $V$ and remove all vertices in $U$,
we have $2d + 4t$ remaining steps to allocate elsewhere.
Therefore, $|Q_{{\textsc{OUT}}}| + |Q_{{\textsc{IN}}}| \leq d + 2t$ as
$Q_{{\textsc{IN}}} \subseteq Y$, $Q_{{\textsc{OUT}}} \subseteq X$, and every
vertex in $Q_{{\textsc{IN}}} \cup Q_{{\textsc{OUT}}}$ must be
touched at least twice (i.e., added and then removed).
Combining those observations, we get:

\begin{center}
$|Q_{{\textsc{OUT}}}| + |Q_{{\textsc{IN}}}| \leq d + 2t$\\
$|Q_{{\textsc{IN}}}| - |Q_{{\textsc{OUT}}}| \leq -d$\\
$|Q_{{\textsc{IN}}}| \leq t$
\end{center}

We have just shown that $G$ has a pair $(Q_{{\textsc{OUT}}}, Q_{{\textsc{IN}}})$
such that $Q_{{\textsc{OUT}}} \subseteq X$, $Q_{{\textsc{IN}}} \subseteq Y$,
$|Q_{{\textsc{IN}}}| \leq t$, $|Q_{{\textsc{OUT}}}| - |Q_{{\textsc{IN}}}| \geq d \geq 0$, and
$N_G(Q_{{\textsc{OUT}}}) = Q_{{\textsc{IN}}}$, as otherwise some edge is not covered.
The remaining condition for $(Q_{{\textsc{OUT}}}, Q_{{\textsc{IN}}})$ to satisfy
is for $G[Q_{{\textsc{OUT}}} \cup Q_{{\textsc{IN}}}]$
to have a matching that saturates $Q_{{\textsc{IN}}}$.
Hall's marriage theorem~\cite{H87} states
that such a saturating matching exists if and only if
for every subset $P$ of $Q_{{\textsc{IN}}}$,
$|P| \leq |N_{G[Q_{{\textsc{OUT}}} \cup Q_{{\textsc{IN}}}]}(P)|$.
By a simple application of Hall's theorem, if
no such matching exists, then there exists a
subgraph $Z$ of $G[Q_{{\textsc{OUT}}} \cup Q_{{\textsc{IN}}}]$
such that $|V(Z) \cap Q_{{\textsc{OUT}}}| < |V(Z) \cap Q_{{\textsc{IN}}}|$.
By~deleting this subgraph
from $Q_{{\textsc{OUT}}} \cup Q_{{\textsc{IN}}}$,
we can get a new pair $(Q_{{\textsc{OUT}}}', Q_{{\textsc{IN}}}')$, which
must still satisfy
$Q'_{{\textsc{OUT}}} \subseteq X$, $Q'_{{\textsc{IN}}} \subseteq Y$,
$|Q'_{{\textsc{IN}}}| \leq t$, $|Q'_{{\textsc{OUT}}}| - |Q'_{{\textsc{IN}}}| \geq d \geq 0$ and
$N_G(Q'_{{\textsc{OUT}}}) = Q'_{{\textsc{IN}}}$, since
we delete more vertices from $Q_{{\textsc{IN}}}$ than
we do from $Q_{{\textsc{OUT}}}$ and
$N_{G[Q_{{\textsc{OUT}}} \cup Q_{{\textsc{IN}}}]}(V(Z) \cap Q_{{\textsc{IN}}}) = V(Z) \cap Q_{{\textsc{OUT}}}$.
Finally, if $(Q_{{\textsc{OUT}}}', Q_{{\textsc{IN}}}')$ does not
have a matching that saturates $Q'_{{\textsc{IN}}}$, we can repeatedly apply the same rule
until we reach a pair that satisfies all the required properties.
Since $|Q_{{\textsc{OUT}}}| \geq |Q_{{\textsc{IN}}}|$, such a pair is guaranteed to exist,
as otherwise every subset $P$ of $Q_{{\textsc{IN}}}$ would satisfy
$|P| > |N_{G[Q_{{\textsc{OUT}}} \cup Q_{{\textsc{IN}}}]}(P)|$ and hence
$|Q_{{\textsc{OUT}}}| < |Q_{{\textsc{IN}}}|$, a contradiction.
\end{proof}

\section{Polynomial-Time Algorithms}\label{section-poly}
In this section, we present a characterization of instances of
the VCR problem solvable in time polynomial in $n$, and apply
this characterization to trees, graphs with no even cycles (as subgraphs) 
and to cactus graphs (with some additional restrictions). We show how to find 
reconfiguration sequences of the shortest
possible length and therefore ignore the parameter $\ell$.
Unless stated otherwise, reconfiguration sequences are
represented as ordered sequences of vertex covers or nodes in
the reconfiguration graph.

\begin{Definition}\label{def-preserv-seq}
Given two vertex covers of $G$, $A$ and $B$, a reconfiguration sequence $\beta$ from $A$ to some
vertex cover $A'$ is a {$c$-bounded prefix} of a reconfiguration sequence $\alpha$ from
$A$ to $B$, if and only if all of the following conditions hold:

\begin{itemize}
\item[(1)] $|A'| \leq |A|$;
\item[(2)] For every node $A''$ in $\beta$, $|A''| \leq |A| + c$;
\item[(3)] For every node $A''$ in $\beta$, $A''$ is
obtained from its predecessor by either the removal or the addition of a single vertex
in the symmetric difference of the predecessor and $B$;
\item[(4)] No vertex is touched more than once in the course of $\beta$.
\end{itemize}

 We write $A \xleftrightarrow{\text{$c, B$}} A'$ when such a
$c$-bounded prefix exists.
\end{Definition}

\begin{Proposition}
Given two vertex covers $S$ and $T$ of $G$, if $G$ has a vertex cover $S'$ such that
$S \xleftrightarrow{\text{$c, T$}} S'$, then $S \xleftrightarrow{\text{$d, T$}} S'$ for all $d > c$.
\end{Proposition}

\begin{Lemma}\label{lemma-preserv-seq}
Given two vertex covers $S$ and $T$ of $G$ and two positive integers $k$ and $c$
such that $|S|, |T| \leq k$, a~reconfiguration sequence $\alpha$
of length $|S_R| + |T_A| = |S \Delta T|$ from $S$ to $T$ exists if:
\begin{itemize}
\item[(1)] $|S| \leq k - c$;
\item[(2)] $|T| \leq k - c$; and
\item[(3)] For any two vertex covers $A$ and $B$ of $G$
such that $|A| \leq k - c$ and $|B| \leq k - c$,
either
$A \xleftrightarrow{\text{$c, B$}} A'$ or $B \xleftrightarrow{\text{$c, A$}} B'$,
where $A'$ and $B'$ are vertex covers of $G$.
\end{itemize}

Moreover, if $c$-bounded prefixes can be found in time polynomial in $n$,
then $\alpha$ can be found in time polynomial in~$n$.
\end{Lemma}

\begin{proof}
We prove the lemma by induction on $|S \Delta T|$. When $|S \Delta T| = 0$,
$S$ is equal to $T$, and the claim holds trivially since $|\alpha| = 0$.

When $|S \Delta T| > 0$, we know that either $S \xleftrightarrow{\text{$c, T$}} S'$
or $T \xleftrightarrow{\text{$c, S$}} T'$.
Without loss of generality, we~assume $S \xleftrightarrow{\text{$c, T$}} S'$
and let $\beta$ denote the $c$-bounded prefix from $S$ to $S'$.
From Definition \ref{def-preserv-seq}, we know that
the size of every node in $\beta$ is no greater than $|S| + c \leq k$.
Therefore, the maximum allowed capacity constraint is never violated.

Since $|S'| \leq |S|$ (Definition \ref{def-preserv-seq}), by the induction hypothesis,
there exists a reconfiguration sequence from $S'$ to $T$ whose
length is $|S' \Delta T|$.
By appending the reconfiguration sequence from $S'$ to $T$ to
the reconfiguration sequence from $S$ to $S'$, we obtain a
reconfiguration sequence $\alpha$ from $S$ to $T$.

To show that $|\alpha| = |S \Delta T|$, it suffices to show that
$|\beta| + |S' \Delta T| = |S \Delta T|$.
We know that no vertex is touched more than once in $\beta$,
and every touched vertex belongs to $S \Delta T$ (Definition \ref{def-preserv-seq}).
We let $H \subseteq S \Delta T$ denote the set
of touched vertices in $\beta$, and we subdivide $H$ into
two sets $H_S = H \cap S = H \cap S_R$ and $H_T = H \cap T = H \cap T_A$.
It follows that $|\beta| = |H_S| + |H_T|$ and
$|S' \Delta T| = |S_R \setminus H_S| + |T_A \setminus H_T|$.
Therefore, $|\beta| + |S' \Delta T| = |H_S| + |H_T| +
|S_R \setminus H_S| + |T_A \setminus H_T| = |S_R| + |T_A| = |S \Delta T|$
as needed.

When $c$-bounded prefixes can be found in time polynomial in $n$,
the proof gives an algorithm for constructing
the full reconfiguration sequence from $S$ to $T$ in time polynomial in $n$.
\end{proof}

\subsection{Trees}
\begin{Theorem}\label{theorem-tree-vc}
{Vertex cover reconfiguration} restricted to trees can be solved in time polynomial in $n$.
\end{Theorem}

\begin{proof}
We let $(G, S, T, k, \ell)$ be an instance of {vertex cover reconfiguration}.
The proof proceeds in two stages. We start by showing
that when $G$ is a tree and $S$ and $T$ are of size at most $k - 1$,
we can always find one-bounded prefixes $S \xleftrightarrow{\text{$1, T$}} S'$ or
$T \xleftrightarrow{\text{$1, S$}} T'$ in time polynomial in $n$. Therefore, we
can apply Lemma \ref{lemma-preserv-seq} with $c = 1$ to find a reconfiguration
sequence of length $|S \Delta T|$ from $S$ to $T$ in time polynomial in $n$.
In the second part of the proof, we show how to handle
the remaining cases where $S$, $T$ or both $S$ and $T$
are of a size greater than $k - 1$.

First, we note that every forest either has a degree-zero or a degree-one vertex.
Hence, trees and forests are one-degenerate graphs.
Since $G$ is a tree, $G[S_R \cup T_A]$ is a forest and is therefore one-degenerate.
To find one-bounded prefixes in $G[S_R \cup T_A]$, it is enough to find a
vertex of degree at most one, which can clearly be done in time polynomial in $n$:
For any two vertex covers $S$ and $T$ of a tree $G$ such that $S, T \leq k - 1$,
we can always find a vertex $v \in S_R \cup T_A$ having degree at most one in $G[S_R \cup T_A]$.
The~existence of $v$ guarantees the existence of
a one-bounded prefix from either $S$ to some vertex
cover $S'$ or from $T$ to some vertex cover $T'$.
When $v \in S_R$ and $|N_{G[S_R \cup T_A]}(v)| = 0$, we have $S \xleftrightarrow{\text{$0, T$}} S'$,
since $S'$ is obtained from $S$ by simply removing $v$.
When $v \in S_R$ and $|N_{G[S_R \cup T_A]}(v)| = 1$, we have $S \xleftrightarrow{\text{$1, T$}} S'$,
since $S'$ is obtained from $S$ by first adding the unique neighbour of $v$ and
then removing $v$. Similar arguments hold when $v \in T_A$.

Therefore, combining Lemma \ref{lemma-preserv-seq} and
the fact that $G[S_R \cup T_A]$ is one-degenerate,
we know that if $|S| \leq k - 1$ and $|T| \leq k - 1$,
a reconfiguration sequence of length $|S_R| + |T_A|$ from $S$ to $T$
can be found in time polynomial in $n$. Furthermore,
since the length of a reconfiguration sequence can never be less than $|S_R| +
|T_A|$, the reconfiguration sequence given by Lemma~\ref{lemma-preserv-seq}
is the shortest path from $S$ to $T$ in the reconfiguration graph.

When $S$ (or $T$) has size $k$ and is minimal, then we have a no-instance,
since neither removing, nor adding a vertex results in a
$k$-vertex cover, and hence, $S$ (or $T$) will be an
isolated node in the reconfiguration graph, with
no path to $T$ (or $S$).

When $S$, $T$ or both $S$ and $T$ are of size $k$ and are
non-minimal, there always exists a reconfiguration sequence from $S$
to $T$, since $S$ and $T$ can be reconfigured to solutions $S'$ and
$T'$, respectively, of size less than $k$, to which
Lemma \ref{lemma-preserv-seq} can be applied.
The only reconfiguration steps from $S$
(or $T$) of size $k$ are to subsets of $S$ of size $k-1$ (or to
subsets of $T$ of size $k-1$); the reconfiguration sequence
obtained from Lemma \ref{lemma-preserv-seq} is thus a shortest path.
Therefore, we can obtain a shortest path from $S$ to $T$ through
a careful selection of $S'$ and $T'$.
There are two cases to consider:

\subparagraph*{Case (1):
 } $|S| = k$, $|T| = k$, $S$ is non-minimal and $T$ is non-minimal.
When both $S$ and $T$ are of size $k$ and are
non-minimal, then each must contain at least one removable vertex.
Hence, by removing such vertices, we can transform $S$ and $T$ into vertex covers $S'$ and
$T'$, respectively, of size $k - 1$.
We let $u$ and $v$ be removable vertices in $S$ and $T$, respectively, and we set
$S' = S \setminus \{u\}$ and $T' = T \setminus \{v\}$.
\begin{enumerate}
\item If $u \in S_R$ and $v \in T_A$, then the length of a
shortest reconfiguration sequence
from $S'$ to $T'$ will be $|S' \Delta T'| = |S \Delta T| - 2$.
Therefore, accounting for the two additional removals,
the length of a shortest path from $S$ to $T$ will be equal to $|S \Delta T|$.

\item If $u \in S_R$ and $v \in C_{ST}$, then the length of a
shortest reconfiguration sequence
from $S'$ to $T'$ will be $|S' \Delta T'| = |S \Delta T| - 1$.
Since $v$ is in $C_{ST}$, it must be removed and added back.
Therefore, the~length of a shortest path from $S$ to $T$ will
be equal to $|S \Delta T| + 2$. The same is true when $u \in C_{ST}$
and $v \in T_A$ or when $u = v$ and $u \in C_{ST}$.

\item Otherwise, when $u \in C_{ST}$, $v \in C_{ST}$ and $u \neq v$, the
length of a shortest path
from $S$ to $T$ will be $|S \Delta T| + 4$, since we
have to touch two vertices in $C_{ST}$ (i.e., two extra
additions and two extra~removals).
\end{enumerate}

\subparagraph*{Case (2): } $|S| = k$, $|T| = k - 1$ and $S$ is non-minimal 
(similar arguments hold for the symmetric case where $|S| = k - 1$, $|T| = k$, and $T$ is non-minimal). 
Since $|T| = k - 1$, we only need to reduce the size of $S$ to $k - 1$
in order to apply Lemma \ref{lemma-preserv-seq}.
Since $S$ is non-minimal, it must contain at least one removable vertex.
We let $u$ be a removable vertex in $S$, and we set $S' = S \setminus \{u\}$.
\begin{enumerate}
\item If $u \in S_R$, then the length of a
shortest reconfiguration sequence
from $S'$ to $T$ will be $|S' \Delta T| = |S \Delta T| - 1$.
Therefore, accounting for the additional removal,
the length of a shortest path from $S$ to $T$ will be equal to $|S \Delta T|$.

\item If $u \in C_{ST}$, then the length of a
shortest reconfiguration sequence
from $S'$ to $T$ will be $|S' \Delta T| = |S \Delta T|$.
Since $v$ is in $C_{ST}$, it must be removed and added back.
Therefore, the length of a shortest path from $S$ to $T$ will
be equal to $|S \Delta T| + 2$.
\end{enumerate}

As there are at most $k^2$ pairs of removable vertices in $S$ and $T$ to check for Case (1),
we can exhaustively try all pairs and choose one that minimizes the length of a
reconfiguration sequence. Similarly, there are at most $k$ removable vertices to check in Case (2).
Consequently, {vertex cover reconfiguration} on trees can be solved in time polynomial in $n$.
\end{proof}

\subsection{Cactus Graphs}
A {cactus graph} $G$~\cite{Brandstadt99} is a connected graph in which every edge belongs to at most one cycle.
We~let ${\cal{C}}(G)$ denote the set of all cycles in $G$.
We say vertex $v \in V(G)$ is a {join vertex} if $v$ belongs to a cycle and $N_G(v) \geq 3$.

The following proposition is a consequence of the fact that 
a maximal matching ${\cal{M}}(G)$ of a cactus graph $G$ can contain an 
edge from each cycle in ${\cal{C}}(G)$. 

\begin{Proposition}\label{obs-cycle-matching}
For a cactus graph $G$, the number of cycles in $G$ is bounded above by the size
of a maximum matching ${\cal{M}}(G)$, i.e., $|{\cal{C}}(G)| \leq |{\cal{M}}(G)|$.
\end{Proposition}

The next proposition is a consequence of the fact that
for any cactus graph $G$, we can obtain a
spanning tree of $G$ by removing a single edge
from every cycle in $G$.

\begin{Proposition}\label{obs-edges-spanning}
For a cactus graph $G$ and $T_G$ a spanning tree of $G$,
the total number of edges in $G$ is equal to the number of
edges in $T_G$ plus the total number of cycles in $G$, i.e.,
$|E(G)| = |E(T_G)| + |{\cal{C}}(G)| = |V(T_G)| - 1 + |{\cal{C}}(G)|$.
\end{Proposition}

Any graph with no even cycles (as subgraphs) is a cactus graph~\cite{Conlon_evencycles}.
For a graph $G$ with no even cycles and any two vertex covers, $S$ and $T$,
of $G$, we know that $G[S_R \cup T_A]$ must be a forest, i.e., a~bipartite
graph with no even cycles (Proposition~\ref{obs-bipartite}).
Proposition~\ref{obs-no-even-cycles} follows from the fact that in the proof
of Theorem \ref{theorem-tree-vc}, the fact that $G$
is a tree is used only to determine that $G[S_R \cup T_A]$ must be a forest.
Therefore, using the same
proof as in Theorem \ref{theorem-tree-vc}, we can show:

\begin{Proposition}\label{obs-no-even-cycles}
{Vertex cover reconfiguration} on graphs with no even cycles can be solved in time polynomial in~$n$.
\end{Proposition}

In the remainder of this section, we generalize Proposition~\ref{obs-no-even-cycles}
to all cactus graphs assuming that the given vertex covers $S$ and $T$ are of size at most $k - 2$. 
To do so, we first show, in Lemmas \ref{lemma-cactus-seq} and \ref{lemma-cactus-preserv-seq}, that~the third condition of Lemma \ref{lemma-preserv-seq} is satisfied
for cactus graphs with $c = 2$.
In Lemma \ref{must-fix-labels}, we show how two-bounded prefixes
can be found in time polynomial in $n$, which leads
to Theorem \ref{theorem-cactus}. We note that a similar result was proven independently by Ito et al.~\cite{ItoNZ16} 
via completely different methods. 

\begin{Lemma}\label{lemma-cactus-seq}
Given two vertex covers $S$ and $T$ of $G$, there
exists a vertex cover $S'$ (or $T'$) of $G$
such that $S \xleftrightarrow{\text{$2, T$}} S'$ (or $T \xleftrightarrow{\text{$2, S$}} T'$)
if one of the following conditions holds:
\begin{itemize}
\item[(1)] $G[S_R \cup T_A]$ has a vertex $v \in S_R$ ($v \in T_A$) such that $|N_{G[S_R \cup T_A]}(v)| \leq 1$; or
\item[(2)] there exists a cycle $Y$ in $G[S_R \cup T_A]$ such that
all vertices in $Y \cap S_R$ ($Y \cap T_A$) have degree exactly two in $G[S_R \cup T_A]$.
\end{itemize}

Moreover, both conditions can be checked in time polynomial in $n$,
and when one of them is true, the corresponding two-bounded prefix
can be found in time polynomial in $n$.
\end{Lemma}

\begin{proof}
First, we note that checking for Condition (1) can be accomplished
in time polynomial in $n$ by simply inspecting the degree of every vertex in $G[S_R \cup T_A]$.
The total number of cycles satisfying condition (2) is linear
in the number of degree-two vertices in $G[S_R \cup T_A]$. Therefore, we can
check for Condition (2) in time polynomial in $n$ by a simple breadth-first
search starting from every degree-two vertex in $G[S_R \cup T_A]$.

If $G[S_R \cup T_A]$ has a vertex $v \in S_R$ of degree zero, we let $S'$
denote the vertex cover obtained by simply removing $v$ from $S$.
It is easy to see that the reconfiguration sequence from $S$ to
$S'$ is a zero-bounded prefix and can be found in time polynomial in $n$.

Similarly, if $G[S_R \cup T_A]$ has a vertex $v \in S_R$
of degree one, we let $S'$ denote the node obtained by
the addition of the single vertex in $N_{G[S_R \cup T_A]}(v)$
followed by the removal of $v$.
The reconfiguration sequence from $S$ to $S'$ is a one-bounded prefix
and can be found in time polynomial in $n$.

For the second case, we let $Y$ be a cycle in $G[S_R \cup T_A]$,
and we partition the vertices of the cycle into
two sets; $Y_S = Y \cap S_R$ and $Y_T = Y \cap T_A$.
Since $G[S_R \cup T_A]$ is bipartite,
we know that $|Y_S| = |Y_T|$. Since all vertices in $Y_S$ have degree
exactly two in $G[S_R \cup T_A]$, it follows
that $N_{G[S_R \cup T_A]}(Y_S) \subseteq Y_T$.
Therefore, a reconfiguration sequence from $S$ to some vertex cover $S'$ that
adds all vertices in $Y_T$ (one by one)
and then removes all vertices in $Y_S$ (one by one) will satisfy
Conditions (1), (3) and (4) from
Definition \ref{def-preserv-seq} for any value of $c$.
For $c = 2$, such a sequence will not satisfy Condition (2) if the cycle has at
least six vertices (i.e., $|Y_T| \geq 3$).
However, using the fact that every vertex in $Y_S$ has degree
exactly two in $G[S_R \cup T_A]$, we can find a reconfiguration sequence from $S$ to
$S'$ in which no vertex cover has a size greater than $|S| + 2$.
To do so, we restrict our attention to $G[Y_S \cup Y_T]$.
Since $Y$ is an even cycle, we~can label all the vertices
of $Y$ in clockwise order from zero to $|Y| - 1$ such that all vertices in $Y_S$ receive
even labels. The reconfiguration sequence from
$S$ to $S'$ starts by adding the two vertices
labelled $1$ and $|Y| - 1$. After doing so, the
vertex labelled 0 is removed.
Next, to remove the vertex labelled 2, we only
need to add the vertex labelled 3.
The same process is repeated for all vertices with even labels up to $|Y| - 4$.
Finally, when we reach the vertex labelled $|Y| - 2$, both of its neighbours
will have already been added, and we can simply remove it.
Hence, we have a two-bounded prefix from $S$ to $S'$, and it is not
hard to see that finding this reconfiguration sequence
can be accomplished in time polynomial in $n$.

When the appropriate assumptions hold, we can show the
symmetric case $T \xleftrightarrow{\text{$2, S$}} T'$ using similar~arguments.
\end{proof}

\begin{Lemma}\label{lemma-cactus-preserv-seq}
If $G$ is a cactus graph and $S$ and $T$ are two vertex covers of $G$,
then there exists a vertex cover $S'$ (or $T'$) of $G$ such that
$S \xleftrightarrow{\text{$2, T$}} S'$ (or $T \xleftrightarrow{\text{$2, S$}} T'$).
\end{Lemma}

\begin{proof}
We assume that $|S_R| \geq |T_A|$, as we can swap the roles of
$S$ and $T$ whenever $|S_R| < |T_A|$.
We observe that every connected component of $G[S_R \cup T_A]$
is a cactus graph since every induced subgraph of a cactus graph is also a cactus graph.
Since we assume $|S_R| \geq |T_A|$, at least one connected component
$X$ of $G[S_R \cup T_A]$ must satisfy $|V(X) \cap S_R| \geq |V(X) \cap T_A|$.

To prove the lemma, we show that if neither condition
of Lemma \ref{lemma-cactus-seq} applies to $X$, it
must be the case that $|V(X) \cap S_R| < |V(X) \cap T_A|$, contradicting our assumption.
To simplify the notation, we assume without loss of generality
that $G[S_R \cup T_A]$ is connected, as we can otherwise set
$G[S_R \cup T_A] = X$.
The~proof proceeds in two steps. First, we show that if Condition (1)
of Lemma \ref{lemma-cactus-seq} is not satisfied, then
$G[S_R \cup T_A]$ must have at least one vertex
$u \in S_R$ of degree at most two in $G[S_R \cup T_A]$.
In the second step, we show that if both Conditions (1) and (2)
of Lemma \ref{lemma-cactus-seq}
are not satisfied, then $|S_R| < |T_A|$,
which completes the proof by contradiction.

Since $G[S_R \cup T_A]$ is a cactus graph, we can apply
Propositions~\ref{obs-cycle-matching} and \ref{obs-edges-spanning} to get:
\vspace{6pt}
\begin{align}\label{equation1}
|E(G[S_R \cup T_A])| &= |S_R| + |T_A| - 1 + |{\cal{C}}(G[S_R \cup T_A])|\notag\\
& \leq |S_R| + |T_A| - 1 + |{\cal{M}}(G[S_R \cup T_A])|
\end{align}

Moreover, since $G[S_R \cup T_A]$ is bipartite
(Proposition~\ref{obs-bipartite}), the size of a
maximum matching in $G[S_R \cup T_A]$ is less than or
equal to $min(|S_R|, |T_A|)$. Therefore:
\begin{align}\label{equation2}
|{\cal{C}}(G[S_R \cup T_A])| \leq |{\cal{M}}(G[S_R \cup T_A])| \leq |S_R|
\end{align}

Combining (\ref{equation1}) and (\ref{equation2}), we get:
\begin{align}\label{equation3}
|E(G[S_R \cup T_A])| &= |S_R| + |T_A| - 1 + {\cal{C}}(G[S_R \cup T_A])\notag\\
& \leq 2|S_R| + |T_A| - 1
\end{align}

If the minimum degree in $G[S_R \cup T_A]$ of any vertex in $S_R$ is three or more, then
$3|S_R| \leq |E(G[S_R \cup T_A])| \leq 2|S_R| + |T_A| - 1$
and thus $|S_R| \leq |T_A| - 1$, contradicting our assumption that $|S_R| \geq |T_A|$.
Hence, $G[S_R \cup T_A]$ must have at least one vertex of degree two in $S_R$.

Next, we show that if $G[S_R \cup T_A]$ has no vertex
$v \in S_R$ such that $|N_{G[S_R \cup T_A]}(v)| \leq 1$ and
no cycle $Y$ such that all vertices in $Y \cap S_R$ have degree
exactly two in $G[S_R \cup T_A]$, then $|S_R| < |T_A|$.
We let $S^x$ denote the set of vertices in $S_R$ having degree $x$ in $G[S_R \cup T_A]$.
Since $G[S_R \cup T_A]$ has no vertex
$v \in S_R$ such that $|N_{G[S_R \cup T_A]}(v)| \leq 1$, we know that $S^2$ cannot be empty.
In addition, since there is no cycle $Y$ in $G[S_R \cup T_A]$ such that all vertices in $Y \cap S_R$
have degree exactly two in $G[S_R \cup T_A]$, any cycle involving
a vertex in $S^2$ must also include a vertex from $\bigcup_{i \geq 3} S^{i}$.
It follows that $\bigcup_{i \geq 3} S^{i}$ is a feedback vertex set (a set whose removal destroys all cycles) of $G[S_R \cup T_A]$,
and $G[S^2 \cup T_A]$ is a forest.

We let $m_s$ denote the maximum degree in $G[S_R \cup T_A]$ of any vertex in $S_R$.
Since each edge in $G[S_R \cup T_A]$ has one endpoint in $S_R$,
\begin{align}\label{equation4}
\sum_{i = 2}^{m_s}{i|S^i|} &\leq |E(G[S_R \cup T_A])|
\end{align}

\noindent
and since each vertex in $S_R$ is in some $S^i$ and using (\ref{equation1}),
we can rewrite (\ref{equation4}) as:
\begin{align}\label{equation5}
\sum_{i = 2}^{m_s}{i|S^i|} &\leq \left( \sum_{i = 2}^{m_s}{|S^i|} \right) + |T_A| - 1 + |{\cal{C}}(G[S_R \cup T_A])|.
\end{align}

To bound $|{\cal{C}}(G[S_R \cup T_A])|$, we note that since no edge
can belong to more than one cycle in a cactus graph,
any vertex $v \in S^x$ can be involved in
at most $\lfloor {x \over 2} \rfloor$ cycles.
Combining this observation with the fact that any cycle
involving a vertex in $S^2$ must also
include a vertex from $\bigcup_{i \geq 3} S^{i}$, we have:
\begin{align}
\sum_{i = 2}^{m_s}{i|S^i|} &\leq \left( \sum_{i = 2}^{m_s}{|S^i|} \right) + |T_A| - 1 + \left( \sum_{i = 3}^{m_s}{{\lfloor{ i \over 2 }\rfloor} |S^i|} \right)\notag\\
& \leq |S^2| + \left( \sum_{i = 3}^{m_s}{(1 + {\lfloor{ i \over 2 }\rfloor})|S^i|} \right) + |T_A| - 1
\end{align}

Finally, by rewriting $\sum_{i = 2}^{m_s}{i|S^i|}$ as $2|S^2| + \sum_{i = 3}^{m_s}{i|S^i|}$
and given that $i - (1 + {\lfloor{ i \over 2 }\rfloor}) \geq 1$ for $i \geq 3$, we~obtain
the desired bound:
\vspace{6pt}
\begin{align}
2|S^2| + \sum_{i = 3}^{m_s}{i|S^i|} &\leq |S^2| + \left( \sum_{i = 3}^{m_s}{(1 + {\lfloor{ i \over 2 }\rfloor})|S^i|} \right) + |T_A| - 1\notag\\
|S^2| + \sum_{i = 3}^{m_s}{i|S^i|} &\leq \left( \sum_{i = 3}^{m_s}{(1 + {\lfloor{ i \over 2 }\rfloor})|S^i|} \right) + |T_A| - 1\notag\\
|S^2| + \sum_{i = 3}^{m_s}{(i - (1 + {\lfloor{ i \over 2 }\rfloor}))|S^i|} &\leq |T_A| - 1\notag\\
|S_R| = \sum_{i = 2}^{m_s}{|S^i|} & \leq |T_A| - 1
\end{align}

This completes the proof.
\end{proof}

\begin{Lemma}\label{must-fix-labels}
If $G$ is a cactus graph and $S$ and $T$ are vertex covers of $G$,
then finding a two-bounded prefix from $S$
to some vertex cover $S'$ (or from $T$ to some vertex cover $T'$) of $G$
can be accomplished in time polynomial in $n$.
\end{Lemma}

\begin{proof}
To find a two-bounded prefix from $S$ to
some vertex cover $S'$ (or from $T$ to some vertex cover $T'$),
we simply need to satisfy one of
the conditions of Lemma \ref{lemma-cactus-seq}, which can both be
checked in time polynomial in $n$.
Since $G[S_R \cup T_A]$ is a cactus graph, we know
from Lemma \ref{lemma-cactus-preserv-seq} that one of them must be~true.
\end{proof}

\begin{Theorem}\label{theorem-cactus}
If $S$ and $T$ are of size at most $k - 2 $, then {vertex cover reconfiguration} on cactus graphs can be solved in time polynomial in $n$.
\end{Theorem}

\begin{proof}
From Lemma \ref{lemma-cactus-preserv-seq}, we know that for
any cactus graph $G$ and two vertex covers $S$ and $T$ of $G$,
then either $S \xleftrightarrow{\text{$2, T$}} S'$ or $T \xleftrightarrow{\text{$2, S$}} T'$,
where $S'$ and $T'$ are some vertex covers of $G$.
In addition, Lemma \ref{must-fix-labels} shows that such two-bounded prefixes
can be found in time polynomial in $n$. By combining these facts, we
can now apply Lemma \ref{lemma-preserv-seq}.
That is, if $|S| \leq k - 2$ and $|T| \leq k - 2$, a reconfiguration sequence of length
$|S_R| + |T_A|$ from $S$ to $T$ can be found in time polynomial in $n$. 
\end{proof}

It remains open whether we can solve the VCR problem in polynomial time on cactus graphs without 
any restrictions on the size of $S$ and $T$. For instance, it is unclear if we 
can always determine (in polynomial time) whether a vertex cover of size $k - 1$ can be 
transformed into a vertex cover of size $k - 2$ and, if so, whether we can find the shortest reconfiguration sequence.

\section{\FPT\ Algorithms}\label{section-fpt}
In this section, we first focus on {vertex cover reconfiguration} on graphs of bounded degree.
We~start by showing that {vertex cover reconfiguration} is \NP-hard on
graphs of degree at most $d$, for~any $d \geq 4$, by proving \NP-hardness on $4$-regular graphs.
The proof is based on the observation that the reconfiguration version of the problem
is at least as hard as the compression version: 

\vspace{8pt}
\noindent
\begin{center}
\begin{tabular}{ll}
\toprule
\multicolumn{2}{l}{{Vertex cover compression}}\\
{\bf Input}: & A graph $G = (V, E)$ and a vertex cover $C$ of $G$ such that $|C| = k \geq 1$ \\
{\bf Parameter}: & $k$\\
{\bf Question}: & Does $G$ have a vertex cover $C'$ of size $k - 1$?\\
\bottomrule
\end{tabular}
\end{center}
\vspace{8pt}

The \NP-hardness result relies on the representation of reconfiguration sequences as edit sequences. 
Next, we give an \FPT
\ algorithm for {vertex cover reconfiguration} on graphs of bounded degree. 
Finally, we show that a host of graph reconfiguration problems definable in first-order logic is fixed-parameter tractable 
on nowhere dense classes of graphs. 

\subsection{Compression via Reconfiguration}
\begin{Theorem}\label{theorem-compression}
{Vertex cover reconfiguration} is at least as hard as {vertex cover compression}.
\end{Theorem}

\begin{proof}
We give a reduction from the latter to the former.
For $(G, C, k)$, an instance of {vertex cover compression}, we
let $V(G) = \{v_1, \ldots, v_n\}$ and form
$G' = (V_G \cup V_A \cup V_B, E_G \cup E_J)$, where
$G'$ consists of the disjoint union of a copy of $G$ and a biclique $K_{k,k}$.
Formally, we have:
\begin{align*}
V_G = \{g_1, \ldots, g_n\}\\
V_A = \{a_1, \ldots, a_k\}\\
V_B = \{b_1, \ldots, b_k\}\\
E_G = \{g_ig_j \mid g_i \in V_G, g_j \in V_G, v_iv_j \in E(G)\}\\
E_J = \{a_ib_j \mid a_i \in V_A, b_j \in V_B, 1 \leq i \leq k, 1 \leq j \leq k\}.\\
\end{align*}

We let $(G', S, T, 3k - 1, 6k - 2)$ be an instance
of {vertex cover reconfiguration}, where
$S = V_A \cup \{g_i \mid v_i \in C\}$ and $T = V_B \cup \{g_i \mid v_i \in C\}$.
Clearly, $|S| = |T| = 2k$ and both $S$ and $T$ are vertex covers of~$G'$.
We claim that $G$ has a vertex cover of size $k - 1$ if and only if
there is a reconfiguration sequence of length $6k - 2$ or less from $S$ to $T$.

Before we can remove any vertex from $V_A$, we need to add all $k$ vertices
from $V_B$. However, $2k + k = 3k > 3k - 1$, which violates the maximum allowed capacity.
Therefore, if there is a reconfiguration sequence from $S$ to $T$, then one of
the vertex covers in the sequence must contain at most $2k - 1$ vertices.
Of those $2k - 1$ vertices, $k$ vertices correspond to the vertices in $V_A$
and cover only the edges in $E_J$. Thus, the remaining $k - 1$ vertices must
be in $V_G$ and should cover all the edges in $E_G$. By our construction
of $G'$, these $k - 1$ vertices correspond to a vertex cover of $G$.

Similarly, if $G$ has a vertex cover $\widehat{C}$ such that $|\widehat{C}| = k - 1$, then the following
reconfiguration sequence transforms $S$ to $T$: add all vertices of
$\widehat{C}$, remove all vertices of $C$, add all vertices of $V_B$, remove all vertices from $V_A$ and
finally add back all vertices of $C$ and remove those of $\widehat{C}$. The length
of this sequence is equal to $6k - 2$ whenever $C \cap \widehat{C} = \emptyset$ and is shorter otherwise.
\end{proof}

\subsection{\NP-Hardness on Four-Regular Graphs}
We are now ready to show that {vertex cover reconfiguration}
remains \NP-hard even if the input
graph is restricted to be four-regular.
We use the same ideas as we did in the previous section.
Since {vertex cover} remains \NP-hard
on four-regular graphs~\cite{GJS74} and any algorithm that solves
the {vertex cover compression} problem can be used to solve
the {vertex cover} problem, we get the desired result.
The main difference here is that we need to
construct a gadget, $W_k$, that is also four-regular.
We describe $W_k$ in terms of several component
subgraphs, each playing a role in forcing
the reconfiguration of vertex covers.

A {$k$-necklace}, $k \geq 4$, is a graph obtained by replacing every edge in
a cycle on $k$ vertices by two vertices and four edges.
For convenience, we refer to every vertex on the original cycle as a {bead}
and every new vertex in the resulting graph as a {sequin}.
The resulting graph has $k$ beads each of degree four and $2k$ sequins each
of degree two. Every two sequins that share the same
neighbourhood in a $k$-necklace are called a {sequin pair}.
We say two beads are {adjacent} whenever they share exactly two
common neighbours. Similarly, we say two sequin pairs are
{adjacent} whenever they share exactly one common neighbour. Every two
adjacent beads (sequin pairs) are {linked} by a sequin pair (bead).

The graph $W_k$ consists of $2k$ copies of a $k$-necklace. We let $U = \{U_1, \ldots U_k\}$
and $L = \{L_1, \ldots L_k\}$ denote the first and second $k$ copies, respectively;
for convenience, we use the terms ``upper'' and ``lower'' to mean ``in $U$'' and ``in $L$'', respectively.
We let $b^u_{i,j}$ and $b^l_{i,j}$ denote the $j$-th beads
of necklace $U_i$ and $L_i$, respectively, where $1 \leq i \leq k$ and $1 \leq j \leq k$.
Beads on each necklace in $W_k$ are numbered
consecutively in ``clockwise order'' from one to $k$.
For every two adjacent beads $b^x_{i,j}$ and $b^x_{i,j + 1}$, where $x \in \{u, l\}$,
we let $p^x_{i,j}$ denote the sequin pair that links both beads.

\begin{figure}[H]
\begin{centering}
\centerline{\includegraphics[scale=0.45]{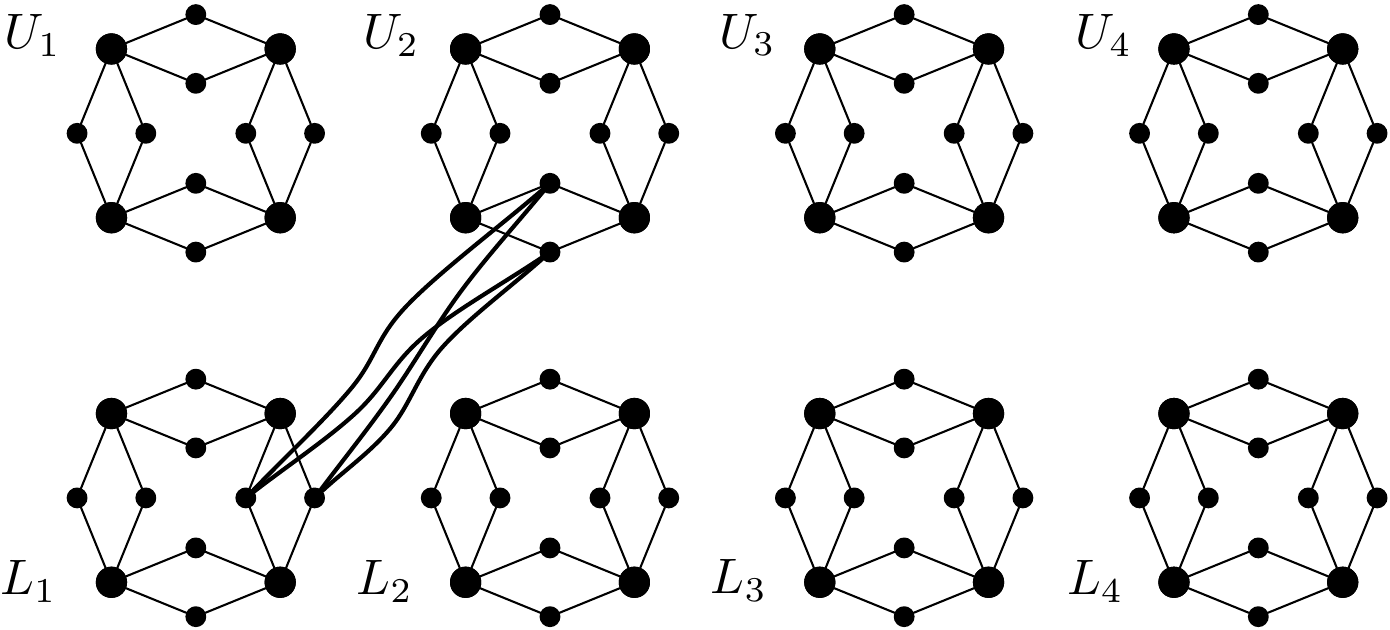}}
\end{centering}
\caption{The graph $W_4$ (the edges of only one of the $k^2$ joining bicliques is shown).}
\label{figure-wk}
\end{figure}

For each sequin pair $p^l_{i,j}$, we add four edges to form
a $K_{2,2}$ (a {joining biclique}) with the
pair $p^u_{j,i}$, for~all $1 \leq i,j \leq k$ (Figure~\ref{figure-wk});
we say that sequin pairs $p^l_{i,j}$ and $p^u_{j,i}$ are {joined}.
All $k^2$ joining bicliques in $W_k$ are vertex disjoint.
The total number of vertices in $W_k$ is
$6k^2$. Every vertex has degree exactly four; every bead
is connected to four sequins
from the same necklace, and every sequin is connected
to two beads from the same necklace
and two other sequins from a different necklace.
We let $S$ be the set containing all upper beads and lower sequins, whereas
$T$ contains all lower beads and upper sequins.
Formally, $S = \{b^u_{i,j} \mid 1 \leq i,j \leq k\} \cup \{v \in p^l_{i,j} \mid 1 \leq i,j \leq k\}$
and $T = \{b^l_{i,j} \mid 1 \leq i,j \leq k\} \cup \{v \in p^u_{i,j} \mid 1 \leq i,j \leq k\}$.
Each set contains $3k^2$ vertices, that is half the vertices in $W_k$.

\begin{Proposition}\label{min-vc}
$S$ and $T$ are minimum vertex covers of $W_k$.
\end{Proposition}

\begin{proof}
We need at least $2k^2$ vertices to cover the edges in the
$k^2$ vertex disjoint joining bicliques contained in $W_k$.
Moreover, any minimal vertex cover $C$ of $W_k$ that includes
a vertex $v$ from a sequin pair $p^x_{i,j} = \{v, w\}$,
where $x \in \{u, l\}$, must also include $w$.
Otherwise, the two beads linking
$p^x_{i,j}$ to its adjacent sequin pairs must be in $C$
to cover the edges incident on $w$, making $v$ removable.
Hence, any~minimal vertex cover $C$ of $W_k$
must include either one or both sequin pairs in a joining biclique.
We~let $x$ denote the number of joining bicliques
from which two sequin pairs are included in $C$.
Similarly, we let $y$ denote the number of joining bicliques
from which only one sequin pair is included in $C$.
Hence, $x + y = k^2$ and $|C| \geq 4x + 2y$.
When $y = 0$, $|C| \geq 4k^2$ and $C$ cannot be a minimum
vertex cover, as $S$ and $T$ are both vertex covers of $W_k$ of size $3k^2$.
When $y \geq 1$, we are left with at least $y$ uncovered edges incident
to the sequin pairs not in $C$.
Those edges must be covered using at least $y$ beads
and, hence, $|C| \geq 4x + 3y$.
If we assume $4x + 3y < 3k^2$, we get a contradiction
since $4x + 4y = 4k^2 < 3k^2 + y$ and $k^2 < y$.
Therefore, $S$ and $T$ must be minimum vertex covers of~$W_k$.
\end{proof}

To prove the next two results, we consider the
representation of reconfiguration sequences as edit sequences. 
Since $S$ is a minimal vertex cover of $W_k$, $\alpha$ cannot
start with a vertex removal. 
Since $V(S, \alpha[1, |\alpha| - 1])$ is a vertex cover of $W_k$,
$|S| = |T|$ and $S$ and $T$ are minimum vertex covers of $W_k$,
$\alpha$~cannot end with a vertex addition. 
Moreover, if $|\alpha| = 6k^2$, then $\alpha$ must touch every vertex in $W_k$ exactly once. 

\begin{Proposition}\label{prop-internal}
Any (valid) edit sequence $\alpha'$ of length $6k^2$ from $S$ to $T$ can be converted into a (valid) edit sequence $\alpha$ from $S$ to $T$ 
such that $|\alpha| = |\alpha'| = 6k^2$, $|V(S, \alpha[1, p])| \leq |V(S, \alpha'[1, p])|$, 
for all $1 \leq p \leq |\alpha|$, and~any two vertices $u$ and $v$ from the same sequin pair in $W_k$ 
are either added in the same maximal addition segment or removed in the same maximal removal segment of $\alpha$. 
Consequently, $\textsf{cap}(\alpha) \leq \textsf{cap}(\alpha')$. 
\end{Proposition}

\begin{proof}
Both vertices in a sequin pair share the same neighbourhood. 
Hence, when $u$ is removed, all~of its neighbours must have been
added, making $v$ also removable.
Moreover, since every vertex is touched exactly once in $\alpha'$,
none of the neighbours of $u$ and $v$ will be touched in $\alpha'$ after the removal of $u$.
Therefore, if $v$ is not removed in the same maximal removal segment as $u$, then we obtain 
$\alpha$ by shifting the removal of $v$, so that it happens immediately after the removal of $u$. 
It is not hard to see that $|\alpha| = |\alpha'| = 6k^2$ and $|V(S, \alpha[1, p])| \leq |V(S, \alpha'[1, p])|$, 
for all $1 \leq p \leq |\alpha|$. 

For the case of additions, if only $u$ is added in some maximal addition segment $\beta$, then none
of its neighbours can yet be removed. Let $\gamma$ be the maximal addition segment in which $v$ is added (which occurs after $\beta$ in $\alpha'$). 
We obtain $\alpha$ by shifting the addition of $u$ from $\beta$ to $\gamma$. 
\end{proof}

\begin{Lemma}\label{lemma-fk}
There exists a function of $k$, $f(k)$, such that $(W_k, S, T, 3k^2 + f(k), \ell)$
is a yes-instance and $(W_k, S, T, 3k^2 + f(k) - 1, \ell)$ is a no-instance
of {vertex cover reconfiguration} for $\ell = 6k^2$.
Moreover, $k - 2\leq f(k) \leq k + 3$.
\end{Lemma}

\begin{proof}
To show that such an $f(k)$ exists, we first prove the $k - 2$ lower bound
by showing that any valid edit sequence $\alpha$ 
of length $6k^2$ from $S$ to $T$ must have some prefix where the number of vertex additions $\#a$ minus 
the number of vertex removals $\#r$ is at least $k - 2$, i.e., $\#a - \#r \geq k - 2$. 
In fact, we will show that the aforementioned property holds for any valid edit sequence $\alpha$ 
of length $6k^2$ in which two vertices from the same sequin pair 
are always added or removed in the same maximal addition or removal segment, respectively. 
Considering only such sequences is sufficient because, from Proposition~\ref{prop-internal}, we know that 
any sequence $\alpha'$ of length $6k^2$ can be transformed into such a sequence $\alpha$ so that 
$|V(S, \alpha[1, p])| \leq |V(S, \alpha'[1, p])|$, for all $1 \leq p \leq |\alpha|$. 
In other words, if $\alpha'$ has no prefix with $\#a - \#r \geq k - 2$ (but $\alpha$ does), then 
$\textsf{cap}(\alpha') < \textsf{cap}(\alpha)$, a contradiction. 

We let position $x$, $1 \leq x \leq |\alpha|$, be the smallest position
such that $\alpha[1,x]$ contains exactly $5k$ vertex removals.
Those $5k$ vertices correspond to a set $S' \subset S$,
as $\alpha$ touches every vertex exactly once.
The~claim is that $\alpha[1,x]$ must contain at least $6k - 2$ vertex additions.
We let $T' \subset T$ denote the set of added vertices in $\alpha[1,x]$.
Since $N_{W_k}(S') \subseteq T'$, we complete the proof of the lower bound
by showing that $|T'| \geq |N_{W_k}(S')| \geq {6 \over 5}|S'| - 2\geq 6k - 2$.
To do so, we show that for any $S' \subset S$ of size $5k$,
$N_{W_k}(S') \subseteq T'$ contains at least ${6 \over 5}|S'| - 2 = 6k - 2$ vertices.

In what follows, we restrict our attention to the bipartite
graph $Z = W_k[S' \cup T']$, and we let $S'$ and $T'$
denote the two partitions of $Z$. We subdivide $S'$ into two sets:
$S'_b$ contains upper beads, and $S'_s$ contains lower sequins.
Since every vertex in $S'_b$ has four neighbours in $T'$ and adjacent
beads share exactly two neighbours, we have $|N_Z(S'_b)| \geq 2|S'_b|$,
and equality occurs whenever $S'_b$ contains $2k$ beads from the same two
upper necklaces. Whenever $S'_b$ contains fewer than $2k$ beads and
$Z[S'_b \cup N_Z(S'_b)]$ consists of $t_b \geq 1$ connected components,
at least one bead from each component (except possibly the first) will be
adjacent to at most one other bead in the same component.
Therefore, $|N_Z(S'_b)| \geq 2|S'_b| + 2(t_b - 1)$.

Proposition~\ref{prop-internal} implies that $T'$ will always
contain both vertices of any sequin pair.
Since we are only considering vertices in $V(\alpha[1,x])$, some
sequins in $S'_s$ might be missing the other sequin in the corresponding pair.
However, all the neighbours of the sequin pair have to be in $T'$, so we assume
without loss of generality that vertices in $S'_s$ can be grouped into sequin pairs.
Every sequin pair in $S'_s$ has four neighbours in $T'$. Adjacent sequin pairs
share exactly one neighbour. Hence, $|N_Z(S'_s)| \geq {3 \over 2}|S'_s|$,
and equality occurs whenever $S'_s$ contains $k$ sequin pairs of a single lower necklace.
Whenever $S'_s$ contains fewer than $k$ sequin pairs and
$Z[S'_s \cup N_Z(S'_s)]$ consists of $t_s \geq 1$ connected components,
at~least one sequin pair from each component will be
adjacent to at most one other sequin pair in the same component.
Therefore, $|N_Z(S'_s)| \geq {3 \over 2}|S'_s| + t_s$.

Combining the previous observations, we know that when
either $S'_b$ or $S'_s$ is empty,
$|N_{Z}(S')| \geq {6 \over 5}|S'|$, as needed.
When both are not empty, we let $I = N_{Z}(S'_b) \cap N_{Z}(S'_s)$.
Hence, $|N_{Z}(S'_b)| + |N_{Z}(S'_s)| - |I|
\geq 2|S'_b| + 2(t_b - 1) + {3 \over 2}|S'_s| + t_s - |I|$,
and we rewrite it as:
\vspace{6pt}
\begin{align}\label{equation-11}
|N_{Z}(S')| + 2 &\geq {100 \over 50}|S'_b| + {75 \over 50}|S'_s| + 2(t_b - 1) + t_s - (|I| - 2)
\end{align}

We now bound the size of $I$. Note that $I$ can only
contain upper sequin pairs joined with sequin pairs in $S'_s$.
As every sequin pair in $S'_s$ has either zero or two neighbours in $I$,
$|S'_s| \geq |I|$. Moreover, for~every two sequin pairs in $S'_s$
having two neighbours in $I$, there must exist at
least one vertex in $S'_b$, which implies $|S'_b| \geq {|I| \over 4}$.
Finally, whenever a sequin pair $p \in S'_s$ has two neighbours in $I$,
then $t_b,t_s \geq 1$, as at least one bead
neighbouring the sequin pair joined with $p$ must be in $S'_b$.
Every other sequin pair $p' \in S'_s$, $p' \neq p$, with
two neighbours in $I$ will force at
least one additional connected component in either
$Z[S'_b \cup N_Z(S'_b)]$ or $Z[S'_s \cup N_Z(S'_s)]$ since $W_k$
contains a single joining biclique between any two necklaces.
Therefore, the total number of
connected components is $t_b + t_s \geq {|I| \over 2}$.
Putting it all together, we get:
\begin{align}\label{equation-22}
{40 \over 50}|S'_b| + {15 \over 50}|S'_s| + 2(t_b - 1) + t_s &\geq {2 \over 10}|I| + {3 \over 10}|I| + {5 \over 10}|I| + t_b - 2\notag\\
&\geq |I| - 2
\end{align}

Combining Equations~\eqref{equation-11} and~\eqref{equation-22}, we get:
\begin{align}\label{equation-33}
|N_{Z}(S')| + 2 &\geq {6 \over 5}|S'| + {40 \over 50}|S'_b| + {15 \over 50}|S'_s| + 2(t_b - 1) + t_s - (|I| - 2) \notag\\
&\geq {6 \over 5}|S'|
\end{align}

Therefore, $V(S, \alpha[1,x])$ is a vertex cover of $W_k$ of size at least
$3k^2 + k - 2$, as needed.

To show the $f(k) \leq k + 3$ upper bound, we show that $(W_k, S, T, 3k^2 + k + 3, 6k^2)$
is a yes-instance by providing an actual reconfiguration sequence:

\begin{itemize}
\item[(1)] Add all $k$ beads in $L_1$.
Since $S$ is a vertex cover of $W_k$, we know that the additional
$k$ beads will result in a vertex cover of size $3k^2 + k$.
\item[(2)] Add both vertices in $p^u_{1,1}$, and remove both vertices in $p^l_{1,1}$.
The removal of both vertices in $p^l_{1,1}$ is possible since we added all their neighbours
in $L_1$ (Step (1)) and $U_1$. The size of a vertex cover reaches $3k^2 + k + 2$ after the
additions and then drops back to $3k^2 + k$.
\item[(3)] Repeat Step (2) for all sequin pairs $p^u_{i,1}$ and $p^l_{1,i}$ for $2 \leq i \leq k$.
The size of a vertex cover is again $3k^2 + k$ once Step (3) is completed.
Step (2) is repeated a total of $k$ times. After every repetition, we have a
vertex cover of $W_k$ since all beads in $L_1$ were added
in Step (1), and the remaining neighbours of each sequin pair
in $U_i$ are added prior to the removals.
\item[(4)] Add both vertices in $p^u_{1,2}$, and remove vertex $b^u_{1,2}$.
\item[(5)] Add $b^l_{2,1}$ and $b^l_{2,2}$. At this point, the size of a vertex cover is $3k^2 + k + 3$.
\item[(6)] Remove both vertices in $p^l_{2,1}$.
\item[(7)] Repeat Steps (4), (5) and (6) until all beads in $L_2$ have been added
and the sequin pairs removed. When we reach the last sequin pair in $L_2$, $b^l_{2,1}$ was already added,
and hence, we gain a surplus of one, which brings the vertex cover size back to $3k^2 + k$.
\item[(8)] Repeat Steps (4) to (7) for every remaining necklace in $L$.
\end{itemize}

Since every vertex in $W_k$ is touched exactly once, we know that $\ell = 6k^2$.
In the course of the described reconfiguration sequence, the maximum size of
any vertex cover is $3k^2 + k + 3$. Hence, $f(k) \leq k + 3$.
This completes the proof.
\end{proof}

It would be interesting to close the gap on $f(k)$, but the existence of such a value is enough to prove
the main theorem of this section.

\begin{Theorem}\label{lemma-singletonA}
{Vertex cover reconfiguration} is \NP-hard on four-regular graphs.
\end{Theorem}

\begin{proof}
We prove the result by a reduction from {vertex cover compression} to
{vertex cover reconfiguration} where the input
graph is restricted to be four-regular in both cases. 
For $(G, C, k)$, an instance of {vertex cover compression}, we
form $G' = (V(G) \cup V(W_k), E(G) \cup E(W_k))$.
We let $(G', S, T, 3k^2 + k + f(k) - 1, 6k^2 + 4k - 2)$ be an instance
of {vertex cover reconfiguration}, where
$S = \{e^u_{i,j} \mid 1 \leq i,j \leq k\} \cup \{p^l_{i,j} \mid 1 \leq i,j \leq k\} \cup C$
and
$T = \{e^l_{i,j} \mid 1 \leq i,j \leq k\} \cup \{p^u_{i,j} \mid 1 \leq i,j \leq k\} \cup C$,
and
$f(k)$ is the value whose existence was shown in Lemma~\ref{lemma-fk}.

Clearly, $|S| = |T| = 3k^2 + k$ and both $S$ and $T$ are vertex covers of $G'$.
We claim that $G$ has a vertex cover of size $k - 1$ if and only if
there is a reconfiguration sequence of length $6k^2 + 4k - 2$ or less from $S$ to $T$.

We know from Lemma~\ref{lemma-fk} that the
reconfiguration of $W_k$ requires at least $f(k)$ available capacity.
However, $3k^2 + k + f(k) > 3k^2 + k + f(k) - 1$, which violates the maximum allowed capacity.
Therefore, if~there is a reconfiguration sequence from $S$ to $T$, then one of
the vertex covers in the sequence must contain at most $3k^2 + k - 1$ vertices.
By Proposition~\ref{min-vc}, we know that $3k^2$ of those $3k^2 + k - 1$ vertices
are needed to cover the edges in $E(W_k)$.
Thus, the remaining $k - 1$ vertices must
be in $V(G)$ and should cover all edges in $E(G)$. By construction
of $G'$, these $k - 1$ vertices correspond to a vertex cover of $G$.

Similarly, if $G$ has a vertex cover $\widehat{C}$ such that $|\widehat{C}| = k - 1$, then the following
reconfiguration sequence transforms $S$ to $T$: add all vertices of
$\widehat{C}$, remove all vertices of $C$, apply the
reconfiguration sequence whose existence was
shown in Lemma~\ref{lemma-fk} to $G'[V(W_k)]$ and finally add back
all vertices of $C$ and remove those of $\widehat{C}$. The length
of this sequence is equal to $6k^2 + 4k - 2$ whenever
$C \cap \widehat{C} = \emptyset$ and is shorter otherwise. 
\end{proof}

\subsection{\FPT\ Algorithm for Graphs of Bounded Degree}
In this section, we prove that {vertex cover reconfiguration} parameterized by $\ell$
is fixed-parameter tractable for graphs of degree at most $d$. 
Our algorithm is randomized and based on a variant of the colour-coding technique~\cite{Alon:1995:COL:210332.210337} that is particularly
useful in designing parameterized algorithms on graphs of bounded
degree. The technique, known in the literature as random separation~\cite{DBLP:conf/iwpec/CaiCC06}, boils
down to a simple, but fruitful observation that in some cases, if we randomly
colour the vertex set of a graph using two colours, the solution or vertices we are
looking for are appropriately coloured with high probability. In our case, we want to make sure 
that the set of touched vertices gets highlighted. We note that our algorithm can easily be 
derandomized using standard techniques~\cite{newpcbook}. 

We start with an instance $(G,S,T,k,\ell)$ of VCR, with $G$ having degree at most $d$. 
Recall that we partition $V(G)$ into the sets $C_{ST} = S \cap T$, $S_R = S \setminus
C_{ST}$, $T_A = T \setminus C_{ST}$, and the independent set $O_{ST} =
V(G) \setminus (S \cup T) = V(G) \setminus (C_{ST} \cup S_R \cup T_A)$. 
We colour independently every vertex of $G$ using one of two colours, say red and blue (denoted by $\mc{R}$ and $\mc{B}$), with probability ${1 \over 2}$. 
We let $\chi: V(G) \rightarrow \{\mc{R}, \mc{B}\}$ denote the resulting random colouring. 
Suppose that $(G,S,T,k,\ell)$ is a yes-instance, and let $\sigma$ denote a reconfiguration sequence 
from $S$ to $T$ of length at most $\ell$. We say that the colouring $\chi$ is {successful} if both of the following conditions hold: 

\begin{itemize}
\item Every vertex in $V(\sigma)$ is coloured red; and 
\item Every vertex in $N_G(V(\sigma))$ is coloured blue. 
\end{itemize}

Observe that $V(\sigma)$ and $N_G(V(\sigma))$ are disjoint. Therefore, the two aforementioned 
conditions are independent. Moreover, since the maximum degree of $G$ is $d$, we have $|V(\sigma)| + |N_G(V(\sigma))| \leq (\ell + 1) d$. 
Consequently, the probability that $\chi$ is successful is at least: 
\begin{align}
\frac{1}{2^{|V(\sigma)| + |N_G(V(\sigma))|}} \geq \frac{1}{2^{(\ell + 1) d}}~~.\notag
\end{align}

Let $V_{\mc{R}}$ denote the set of vertices coloured red and $V_{\mc{B}}$ denote the set 
of vertices coloured blue. Moreover, we let $C_1$, $\dots$, $C_q$ denote the set of connected components of $G[V_{\mc{R}}]$. 
The main observation now is the following: 

\begin{Proposition}\label{prop-colouring}
If $\chi$ is successful, then $N_G(V(\sigma)) \subseteq C_{ST}$, 
$V(\sigma)$ has a non-empty intersection with at most $\ell$ connected components of $G[V_{\mc{R}}]$, 
and each one of those components consists of at most $\ell$ vertices. 
\end{Proposition}

\begin{proof} 
The fact that $N_G(V(\sigma)) \subseteq C_{ST}$ follows from the observation that every vertex in $V(\sigma)$ must be 
added or removed at least once and no vertex in $N_G(V(\sigma))$ is ever added or removed. 
In other words, if $v \in V(\sigma)$ is removed, then all of its untouched neighbours must be in $C_{ST}$. Similarly, if $v \in V(\sigma)$ is added, then prior 
to being added, all of its untouched neighbours must be in $C_{ST}$. 

Since $|V(\sigma)| \leq \ell$, we know that $G[V(\sigma) \cup N_G(V(\sigma))]$ consists of at most $\ell$ connected components 
(each of size at most $(\ell + 1)d$) and $G[V(\sigma)]$ consists of at most $\ell$ components (each of size at most $\ell$). 
Let~$C$ be a connected component of $G[V_{\mc{R}}]$ such that $|V(C)| > \ell$. We claim that we can safely ignore (and hence delete)
this component when $\chi$ is successful. Suppose to the contrary that $V(\sigma) \cap V(C) = Q \neq \emptyset$. 
Since $\chi$ is successful, it must be the case that every vertex in $N_G(Q)$ is coloured blue. However, we know that there exists 
at least one vertex in $N_G(Q)$ that is coloured red (since $C$ is a connected component of $G[V_{\mc{R}}]$ and all vertices in $C$ are coloured red). 
As we have obtained a contradiction, we can conclude that when $\chi$ is successful, $V(\sigma)$ can intersect at most $\ell$ connected components of 
$G[V_{\mc{R}}]$, and none of those components can be of a size greater than $\ell$, as claimed.
\end{proof}

Given Proposition~\ref{prop-colouring}, we can safely assume that every connected 
component of $G[V_{\mc{R}}]$ consists of at most $\ell$ vertices (as the remaining components can be ignored when $\chi$ is successful). 
For simplicity, let~us first assume that $G[V(\sigma)]$ is connected. 
Thus, if $\chi$ is successful, then there exists a single component in $G[V_{\mc{R}}]$, say $C^\star$, 
such that $V(\sigma) \subseteq V(C^\star)$, $|V(C^\star)| \leq \ell$ and $S_R \cup T_A \subseteq V(C^\star)$. 
Therefore, we can simply enumerate all possible sequences of length at most $\ell$ and make sure 
that at least one of them is the required reconfiguration sequence from $S$ to $T$. 
This brute-force testing can be accomplished in time $2^{\Oh(\ell \log \ell)} \cdot n^{\Oh(1)}$. 

Let us now consider the general case when $G[V(\sigma)]$ is not necessarily connected. 
We say a component $C$ of $G[V_{\mc{R}}]$ is {important} if $V(C) \cap (S_R \cup T_A) \neq \emptyset$. 
There are at most $\ell$ important components. Hence, we only need to bound the number of unimportant components. 
To that end, we partition the unimportant components of $G[V_{\mc{R}}]$ into equivalence classes with 
respect to the following relation $\simeq$: 
$$
C \simeq C' \quad \Leftrightarrow \quad C~\text{is isomorphic to}~C'.
$$

\begin{Proposition}\label{cl:num-classes2}
The total number of graphs with at most $\ell$ vertices is at most $2^{\Oh(\ell^2)}$, and therefore, 
the equivalence relation $\simeq$ has at most $2^{\Oh(\ell^2)}$ equivalence classes. 
\end{Proposition}

Assume that some equivalence class contains more than $\ell$ unimportant components. 
We claim that retaining only $\ell$ of them is enough. 
To see why, it is enough to note that $V(\sigma)$ intersects with at most $\ell$ of those components; they are all isomorphic; 
and the neighbours of any such component are contained in $C_{ST}$. Putting it all together, we know that we have at most 
$2^{\Oh(\ell^2)}$ equivalence classes, each with at most $\ell$ components, and each component is of size at most $\ell$. 
Hence, we can guess the sequence from $S$ to $T$ in time $2^{\Oh(\ell^3 \log \ell)} \cdot n^{\Oh(1)}$ 
(testing whether two graphs with $\ell$ vertices are isomorphic can be accomplished naively in time $2^{\ell \log \ell}$). 

We have proven that the probability that $\chi$ is successful is at least $2^{-(\ell + 1)d}$. 
Hence, to obtain a Monte Carlo algorithm with false negatives, we repeat the
above procedure $2^{(\ell + 1)d}$ times and obtain the following result:

\begin{Theorem}
There exists a one-sided error Monte Carlo algorithm with false negatives that
solves the {vertex cover reconfiguration} problem on graphs of degree at most $d$ in time $2^{(\ell + 1)d} \cdot 2^{\Oh(\ell^3 \log \ell)} \cdot n^{\Oh(1)}$. 
\end{Theorem}

\subsection{\FPT\ Algorithm for Nowhere Dense Graphs}

\renewcommand{\phi}{\varphi}

In this section, we present a more general result, showing that 
the reconfiguration variant of every first-order definable optimization problem 
parameterized by $\ell$ is fixed-parameter tractable on every fixed 
nowhere dense class of graphs. 

Let us quickly recall the necessary definitions from logic. 
For our purpose, it suffices to consider
first-order logic over the vocabulary of coloured graphs. 
We refer to the textbook~\cite{hodges1993model} for 
extensive background on logic. 
 
Let $A,B$ be two unary relation symbols and $E$ a binary
relation symbol. We call $\{E,A,B\}$ the {vocabulary}
of graphs with two colours $A$ and $B$. {First-order formulas} over the vocabulary of coloured graphs are
formed from atomic formulas~$x=y$, $E(x,y)$, $A(x)$ and $B(x)$, where $x,y$ are variables (we~assume that we have an infinite supply of variables), by the usual Boolean
connectives~$\neg$~(negation),~$\wedge$~(conjunction) and~$\vee$ (disjunction)
and existential and universal quantification~$\exists x,\forall x$,
respectively. The~{free variables} of a formula are those not in the scope of a
quantifier, and we write~$\phi(x_1,\ldots,x_k)$ to indicate that the free
variables of the formula~$\phi$ are among $x_1,\ldots,x_k$.
To define the semantics, we inductively define a satisfaction
relation~$\models$. Let $G$ be a graph and $A,B\subseteq V(G)$. For simplicity,
we do not distinguish between $A$ as a symbol and $A$ as a set. For~a formula~$\phi(x_1,\ldots,x_k)$, and
$v_1,\ldots,v_k\in V(G)$, $G\models\phi(v_1,\ldots,v_k)$
means that~$G$ satisfies~$\phi$ if the free variables~$x_1,\ldots,x_k$
are interpreted by~$v_1,\ldots,v_k$. If
$\phi(x_1,x_2)=E(x_1,x_2)$ is atomic, then $G\models\phi(v_1,v_2)$
if~$(v_1,v_2)\in E(G)$. Similarly, if $\phi(x)=A(x)$, then $G\models\phi(v)$
if~$v\in A$. The meanings of the equality symbol, the
Boolean connectives and the quantifiers are the usual ones. 

Let $\phi(X)$ be a first-order formula that has a free
set variable $X$. For example, 
the vertex cover problem is defined by the formula:
\[\phi(X)=\forall x\forall y (\neg E(x,y) \vee X(x) \vee X(y)).\] 

As another example, we can define dominating sets by the formula:
\[\phi(X)=\forall x (X(x) \vee \exists y(X(y)\wedge E(x,y))\]
and independent sets by the formula:
\[\phi(X) = \forall x\forall y((X(x)\wedge X(y)\wedge x\neq y)
\rightarrow \neg E(x,y)).\]

Naturally, we can define a
reconfiguration variant for each such formula $\phi$. Given two
sets $S,T\subseteq V(G)$, we ask for a sequence $S_1,\ldots, S_t$, 
$S_1=S$ and $S_t=T$ such that $G\models\phi(S_i)$ for all 
intermediate configurations $S_i$. We call
the corresponding decision problem $\phi$-{reconfiguration}, and we refer to a 
solution of a problem instance as a {$\phi$-reconfiguration sequence}. 

Nowhere dense graph classes were introduced by Ne\v{s}et\v{r}il and
Ossona de Mendez~\cite{nevsetvril2011nowhere} as a very general
model of uniform sparseness in graphs. We refer to the 
textbook~\cite{nevsetvril2012sparsity} for the formal definition of
the notion of nowhere denseness and for more background on its theory. 
For our purpose, it is sufficient to note that most familiar classes of sparse
graphs are nowhere dense, e.g.,\ every proper minor closed class and
every class of bounded degree is nowhere dense. We will prove the following~theorem. 

\begin{Theorem}
Let $\phi(X)$ be a first-order formula over the vocabulary of graphs 
with a free set variable; let $\mathcal{C}$ be a nowhere dense class of graphs; 
let $\epsilon>0$ be a real; and let $\ell\geq 1$ be an integer. Then, there exists 
a constant $f(|\phi|,\ell,\epsilon)$ and an algorithm that, given an $n$-vertex
graph $G\in\mathcal{C}$ and two sets $S,T\subseteq V(G)$ with 
$G\models\phi(S)$, $G\models \phi(T)$, decides whether there exists
a $\phi$-reconfiguration sequence of length at most $\ell$ in time 
$f(|\phi|,\ell,\epsilon)\cdot n^{1+\epsilon}$. 
\end{Theorem}

\begin{proof}
Our proof is based on a result of Grohe, Kreutzer and Siebertz~\cite{grohe2017deciding}, 
which states that for every first-order 
formula $\psi$ (without free variables), every nowhere dense
class $\mathcal{C}$ of graphs and every real \mbox{$\epsilon>0$,} 
there exists a constant $f(|\psi|,\epsilon)$, such that given an $n$-vertex 
graph $G\in\mathcal{C}$, 
one can decide in time $f(|\psi|,\epsilon)\cdot n^{1+\epsilon}$
whether $\psi$ holds in $G$. 

In order to approach the $\phi$-reconfiguration problem, 
we want to write a formula~$\psi$ over the vocabulary of graphs
with two colours $S$ and $T$ without free variables, which expresses 
the existence of a $\phi$-reconfiguration sequence of length at most~$\ell$. 
We will guarantee that the length of $\psi$ is bounded by a function 
depending only on $\ell$ (and on $\phi$, though only as a fixed constant). 
Then, by fixing any $\epsilon>0$ and using 
the result of~\cite{grohe2017deciding}, we conclude
that $\phi$-{Reconfiguration} is fixed-parameter tractable parameterized
by $\ell$ on every nowhere dense class $\mathcal{C}$. 

The formula $\psi$ simply states the existence of a sequence
of $\ell$ elements that will be added or removed in the course 
of the reconfiguration. 
For each initial sequence of length $i=1,\ldots, \ell$ 
of these~$\ell$ guesses, we state in $\psi$ that the formula $\phi(X)$ is
satisfied for the set $S$ modified according to the first~$i$ operations. Finally, we state that the reconfiguration leads to 
the set $T$. The precise formula is cumbersome to write; however, 
we expect that the reader is convinced that we can express 
the desired statement in first-order logic once we have stated
how to handle a single addition and removal of a vertex. To state
that a vertex is added to the set $S$, we write the formula:
\[\exists x_1 \big(\phi'(S,x_1)\big),\] where $\phi'(X)$ is
obtained from $\phi(X)$ by replacing every atom $X(x)$ 
for a variable $x$ by the formula $X(x)\vee x=x_1$. 
If we now want to remove a vertex, we extend the formula
to the formula: \[\exists x_2\exists x_1\Big(\big((x_1\neq x_2)\rightarrow
\phi''(S,x_1,x_2)\big)\wedge \big((x_1=x_2)\rightarrow \phi'''(S,x_2)\big)\Big),\] 
where $\phi''(X,x_1,x_2)$ is the formula obtained from $\phi(X)$ by 
replacing every atom $X(x)$ by the formula $(X(x)\vee x=x_1)
\wedge x\neq x_2$, and $\phi'''(X,x_2)$ is obtained from 
$\phi(X)$ by replacing $X(x)$ by $X(x)\wedge x\neq x_2$. 
Similarly, we can handle any sequence of additions and removals
of vertices of length at most $\ell$ and form a big disjunction
over all such sequences to obtain the final formula $\psi$. 
\end{proof}

\section{Conclusions}
To the best of our knowledge, our results constitute the first in-depth
study of the VCR problem parameterized
by the length of a reconfiguration sequence. We showed that even though
the {vertex cover} problem is solvable in polynomial time on bipartite
graphs, VCR remains \WO-hard.
On the tractable side, we showed that VCR 
is solvable in polynomial time for trees, as well as graphs with no even cycles 
and is fixed-parameter tractable for graphs of bounded degree.
It remains open whether we can solve VCR in polynomial time on cactus graphs without 
any restrictions on the size of $S$ and $T$. 
Finally, we believe that the techniques used in both our hardness proofs
and positive results can be extended to cover a host of graph deletion
problems defined in terms of hereditary graph properties~\cite{MNRSS13}.
It also remains to be seen whether our \FPT\ results can be extended
to a larger class of sparse graphs, e.g., biclique-free graphs. 

\subparagraph*{Acknowledgments.}
The authors would like to thank Daniel Lokshtanov for fruitful discussions that greatly helped improve the presentation of some of the results.

\bibliographystyle{plain}	
\bibliography{vcrecon_full3_refs}

\begin{thebibliography}{10}

\bibitem{ACFLSS09}
Faisal~N. Abu-Khzam, Rebecca~L. Collins, Michael~R. Fellows, Michael~A.
  Langston, W.~Henry Suters, and Christopher~T. Symons.
\newblock Kernelization algorithms for the vertex cover problem: Theory and
  experiments.
\newblock In {\em Proc. of the Sixth Workshop on Algorithm Engineering and
  Experiments}, pages 62--69, 2004.

\bibitem{Alon:1995:COL:210332.210337}
Noga Alon, Raphael Yuster, and Uri Zwick.
\newblock Color-coding.
\newblock {\em J. ACM}, 42(4):844--856, 1995.

\bibitem{B12}
Paul Bonsma.
\newblock The complexity of rerouting shortest paths.
\newblock In {\em Proc. of Mathematical Foundations of Computer Science}, pages
  222--233, 2012.

\bibitem{Brandstadt99}
Andreas Brandst\"{a}dt, Van~Bang Le, and Jeremy~P. Spinrad.
\newblock {\em Graph Classes: A Survey}.
\newblock Society for Industrial and Applied Mathematics, Philadelphia, PA,
  USA, 1999.

\bibitem{DBLP:conf/iwpec/CaiCC06}
Leizhen Cai, Siu~Man Chan, and Siu~On Chan.
\newblock Random separation: {A} new method for solving fixed-cardinality
  optimization problems.
\newblock In {\em Parameterized and Exact Computation, Second International
  Workshop, {IWPEC} 2006, Z{\"{u}}rich, Switzerland, September 13-15, 2006,
  Proceedings}, pages 239--250, 2006.

\bibitem{CVJ08}
Luis Cereceda, Jan van~den Heuvel, and Matthew Johnson.
\newblock Connectedness of the graph of vertex-colourings.
\newblock {\em Discrete Mathematics}, 308(56):913--919, 2008.

\bibitem{CFJ04}
Benny Chor, Michael~R. Fellows, and David Juedes.
\newblock Linear kernels in linear time, or how to save $k$ colors in
  ${O}(n^2)$ steps.
\newblock In {\em Proc. of the 30th International Conference on Graph-Theoretic
  Concepts in Computer Science}, pages 257--269, 2004.

\bibitem{Conlon_evencycles}
Joseph~G. Conlon.
\newblock Even cycles in graphs.
\newblock {\em Journal of Graph Theory}, 45(3), 2004.

\bibitem{newpcbook}
Marek Cygan, Fedor~V. Fomin, Lukasz Kowalik, Daniel Lokshtanov, D{\'{a}}niel
  Marx, Marcin Pilipczuk, Michal Pilipczuk, and Saket Saurabh.
\newblock {\em Parameterized Algorithms}.
\newblock Springer, 2015.

\bibitem{D05}
Reinhard Diestel.
\newblock {\em {Graph Theory}}.
\newblock Springer-Verlag, Electronic Edition, 2005.

\bibitem{DF97}
Rod~G. Downey and Michael~R. Fellows.
\newblock {\em Parameterized complexity}.
\newblock Springer-Verlag, New York, 1997.

\bibitem{DowneyF13}
Rodney~G. Downey and Michael~R. Fellows.
\newblock {\em Fundamentals of Parameterized Complexity}.
\newblock Texts in Computer Science. Springer, 2013.

\bibitem{FHHH11}
Gerd Fricke, Sandra~Mitchell Hedetniemi, Stephen~T. Hedetniemi, and Kevin~R.
  Hutson.
\newblock $\gamma$-{G}raphs of {G}raphs.
\newblock {\em Discussiones Mathematicae Graph Theory}, 31(3):517--531, 2011.

\bibitem{GJS74}
M.~R. Garey, David~S. Johnson, and Larry~J. Stockmeyer.
\newblock Some simplified {NP}-complete graph problems.
\newblock {\em Theor. Comput. Sci.}, 1(3):237--267, 1976.

\bibitem{GKMP09}
Parikshit Gopalan, Phokion~G. Kolaitis, Elitza~N. Maneva, and Christos~H.
  Papadimitriou.
\newblock The connectivity of boolean satisfiability: computational and
  structural dichotomies.
\newblock {\em SIAM J. Comput.}, 38(6):2330--2355, 2009.

\bibitem{grohe2017deciding}
Martin Grohe, Stephan Kreutzer, and Sebastian Siebertz.
\newblock Deciding first-order properties of nowhere dense graphs.
\newblock {\em Journal of the ACM (JACM)}, 64(3):17, 2017.

\bibitem{H87}
Philip Hall.
\newblock On representatives of subsets.
\newblock In {\em Classic Papers in Combinatorics}, pages 58--62. Birkhäuser
  Boston, 1987.

\bibitem{hodges1993model}
Wilfrid Hodges.
\newblock {\em Model theory}, volume~42.
\newblock Cambridge University Press, 1993.

\bibitem{IDHPSUU11}
Takehiro Ito, Erik~D. Demaine, Nicholas J.~A. Harvey, Christos~H.
  Papadimitriou, Martha Sideri, Ryuhei Uehara, and Yushi Uno.
\newblock On the complexity of reconfiguration problems.
\newblock {\em Theor. Comput. Sci.}, 412(12-14):1054--1065, 2011.

\bibitem{IKD12}
Takehiro Ito, Marcin Kami\'{n}ski, and Erik~D. Demaine.
\newblock Reconfiguration of list edge-colorings in a graph.
\newblock {\em Discrete Applied Mathematics}, 160(15):2199--2207, 2012.

\bibitem{IKOZ12}
Takehiro Ito, Kazuto Kawamura, Hirotaka Ono, and Xiao Zhou.
\newblock Reconfiguration of list {L(2, 1)}-labelings in a graph.
\newblock In {\em Proc. of the 23rd Annual International Symposium on
  Algorithms and Computation}, pages 34--43, 2012.

\bibitem{ItoNZ16}
Takehiro Ito, Hiroyuki Nooka, and Xiao Zhou.
\newblock Reconfiguration of vertex covers in a graph.
\newblock {\em {IEICE} Transactions}, 99-D(3):598--606, 2016.

\bibitem{KMM11}
Marcin Kami\'{n}ski, Paul Medvedev, and Martin Milani\v{c}.
\newblock Shortest paths between shortest paths.
\newblock {\em Theor. Comput. Sci.}, 412(39):5205--5210, 2011.

\bibitem{KMM12}
Marcin Kami\'{n}ski, Paul Medvedev, and Martin Milani\v{c}.
\newblock Complexity of independent set reconfigurability problems.
\newblock {\em Theor. Comput. Sci.}, 439:9--15, 2012.

\bibitem{DBLP:conf/isaac/MouawadNR14}
Amer~E. Mouawad, Naomi Nishimura, and Venkatesh Raman.
\newblock Vertex cover reconfiguration and beyond.
\newblock In {\em Algorithms and Computation - 25th International Symposium,
  {ISAAC} 2014, Jeonju, Korea, December 15-17, 2014, Proceedings}, pages
  452--463, 2014.

\bibitem{MNRSS13}
Amer~E. Mouawad, Naomi Nishimura, Venkatesh Raman, Narges Simjour, and Akira
  Suzuki.
\newblock On the parameterized complexity of reconfiguration problems.
\newblock {\em Algorithmica}, 78(1):274--297, 2017.

\bibitem{nevsetvril2012sparsity}
Jaroslav Ne{\v{s}}et{\v{r}}il and P~Ossona De~Mendez.
\newblock Sparsity.
\newblock {\em Algorithms and Combinatorics}, 28, 2012.

\bibitem{nevsetvril2011nowhere}
Jaroslav Ne{\v{s}}et{\v{r}}il and Patrice~Ossona de~Mendez.
\newblock On nowhere dense graphs.
\newblock {\em European Journal of Combinatorics}, 32(4):600--617, 2011.

\bibitem{Wrochna18}
Marcin Wrochna.
\newblock Reconfiguration in bounded bandwidth and tree-depth.
\newblock {\em J. Comput. Syst. Sci.}, 93:1--10, 2018.

\end{thebibliography}

\end{document}